\documentclass[ecta,nameyear]{econsocart}
\makeatletter
\let\ps@copyright\ps@plain
\makeatother


\usepackage[colorlinks,citecolor=blue,linkcolor=blue,urlcolor=blue,pagebackref]{hyperref}

\startlocaldefs

\usepackage{graphicx, subfigure}
\usepackage{rotating, booktabs}
\usepackage{rotfloat}
\usepackage{tikz} 
\usepackage{setspace}
\newcommand{\sym}[1]{^{#1}}

\newtheorem*{theorem}{General identification logic}
\newtheorem*{theorem2}{Spatial control group}
\theoremstyle{definition}

\usepackage{xparse}
\usepackage{natbib}
\usepackage{xcolor}

\endlocaldefs
\NewDocumentCommand{\indicator}{}{\text{\usefont{U}{stix2bb}{m}{n}1}}
\usepackage{comment}
\begin{document}

\begin{frontmatter}

\title{Identifying Time Preference Shocks: Stated and Revealed Preferences After an Italian Earthquake}
\runtitle{Identifying Time Preference Shocks}

\begin{aug}
\author[add1]{\fnms{Krisztina}~\snm{Molnar}\ead[label=e1]{krisztina.molnar@nhh.no}}
\author[add2]{\fnms{Concetta}~\snm{Rondinelli}\ead[label=e2]{concetta.rondinelli@bancaditalia.it}}
\author[add3]{\fnms{Po Yin}~\snm{Wong}\ead[label=e3]{po.wong@qmul.ac.uk}}
\address[add1]{%
\orgdiv{Department of Economics},
\orgname{Norwegian School of Economics}}

\address[add2]{%
\orgdiv{Directorate General for Economics, Statistics and Research},
\orgname{Bank of Italy}}

\address[add3]{%
\orgdiv{Queen Mary Business School},
\orgname{Queen Mary University of London}}
\end{aug}

\begin{funding}
This version: November 2026. We thank Ludovica Gazze, Giovanni Marin,  Ilan Noy, Pedro Bordalo, Jonna Olsson, Jon Steinsson and participants at conferences (CEPR, EEA-ESEM, RES, SES, OMG), Ana Carolina Pessoa for creation of the map, and Karlygash Kuralbayeva for her contributions to the early stages of this research. Remaining errors are our own. The views expressed herein are those of the authors and should not be attributed to the Bank of Italy.
\end{funding}

\noindent%
\begin{abstract}
Identifying time preference shocks is challenging: shocks large enough to shift survey-elicited stated impatience also tighten financial resources, confounding the interpretation of stated time preference. Jointly examining stated and revealed preferences resolves this identification problem: impatience shocks and reductions in financial resources generate opposite consumption responses, even though both raise stated impatience. Applying our identification framework to the 2009 L'Aquila earthquake, we combine panel data with spatially granular seismic and damage data in a difference-in-differences design that compares undamaged households that experienced the earthquake with a spatial control group at similar distances from the epicenter. Treated households report greater impatience and increase nondurable consumption by 11 percent relative to controls.
Effects are consistent with a psychological-stress channel, concentrated among the most vulnerable to stress, and persist as households deplete their liquid wealth buffer over time. 
\end{abstract}

\begin{keyword}
\kwd{Time preference shocks}
\kwd{identification}
\kwd{hypothetical questions}
\kwd{consumption}
\kwd{natural disaster}
\kwd{Italy}
\end{keyword}

\begin{keyword}[class=JEL]
\kwd{D15}
\kwd{D91}
\kwd{E21}
\kwd{Q54}
\kwd{C23}
\end{keyword}

\end{frontmatter}

 \section{Introduction}\label{s1}

\begin{quote}\flushright \textit{You ask people questions, but you want to be sure that you interpret correctly what they are saying.} \\(Orazio Attanasio\footnote{16 March 2023 in the ``Coffee with...''  \href{https://www.nhh.no/en/research-centres/fair/news/2023/coffee-with-orazio-attanasio/}{\textcolor{blue}{interview of the FAIR Insight Team}}})
\end{quote}

Wars, floods, earthquakes, and hurricanes have all reportedly changed how people value the future. A soldier who survived Burundi's civil war turns out to discount tomorrow more heavily than one who did not experience the war \citep{Voors_etal12AER_burundiviolence}; a Sri Lankan fisherman swept up by the Indian Ocean tsunami emerges, puzzlingly, more patient than before \citep{disasters_indian_ocean_earthquake_callen15}. Economists have documented shifts in \textit{stated time preferences} - measured through survey questions on hypothetical intertemporal choices - across a wide range of natural disasters and conflict episodes. But interpreting what people say about their preferences turns out to be harder than it looks. 
Wars and natural disasters also destroy financial resources through lower incomes or severed credit access, which \cite{carvalho2016poverty, Dean21time-cc} showed confounds stated time preferences.\footnote{Much of the earlier literature acknowledges the difficulty of separating genuine preference shifts from changes driven by economic hardship; therefore, interpret stated preference changes as contextual behavioral responses   \citep{CallenAER14_afghanistan, disasters_11floodCambodia_sommarat_2016, disasters_tsunami_thailand_cassar17,KUROISHI2024}.}
Distinguishing a genuine change in the rate of time preference from a rational response to lower financial resources is an identification problem the literature has not yet solved.\footnote{\cite{carvalho2016poverty, Dean21time-cc} show this bias operates specifically through monetary tradeoffs: individuals with lower financial resources appear more impatient when choosing between money-now versus money-later, but not when choosing between real-effort tasks at different times.  On how financial resources distort responses to hypothetical scenarios, see also \citet{guiso2008trusting-stockmkt, Haushofer2014PsychologyPoverty, ClausThustrupKreiner_wealyh_beta}. The macro literature makes the same point on the revealed-preference side: constraints can mimic impatience in consumption behavior as well as in survey answers \citep{McKayNakamuraSteinsson2016}.} If stated impatience reflects lack of financial resources, transfers can reverse it; if latent time preferences have shifted, the change is structural, and transfers do not reverse it. 

We make progress in identifying time preference shocks by proposing a general sign identification framework that uses changes in both stated and revealed time preferences (intertemporal choice behavior). Stated time preferences might change similarly in response to time preference shocks and changes in financial resources \citep{ambrus2015, Epper2015, carvalho2016poverty, Dean21time-cc}, but consumption-saving responses differ.
An impatience shock increases current consumption, as households with sufficient liquidity front-load expenditure, while lower financial resources compress current consumption, or reduce savings if households smooth consumption \citep{Carroll1997}.
This simple logic parallels a classic sign-identification strategy in macroeconomics, imposing logical, theory-based directional constraints, or ``signs'', on how prices and quantities respond to unexpected demand and supply shocks \citep{manski1997monotone, uhlig2005effects}. Recent literature promotes analyzing stated and revealed preferences jointly (summarized in  \citet{Almas2024}). While earlier research builds on the auxiliary assumption of a stable mapping between stated and revealed preferences, we instead treat the divergence between stated and revealed preferences as the identifying signal. \footnote{\citet{bernheimBjorkegrenNaeckerPollman2021causal} and \citet{meango2025combiningstatedrevealedpreferences} both develop new identification frameworks based on such a stable mapping, the former to estimate treatment effects, the latter to recover latent heterogeneity.}

We apply this framework to an exceptional dataset, where stated time preferences, stated risk preferences, and household choices can be observed jointly in a panel, bracketing the 2009 L'Aquila earthquake in Italy. The time preference measure is a hitherto unexploited feature of the Bank of Italy's Survey of Household Income and Wealth (SHIW), elicited in two waves before and two waves after the earthquake. SHIW is a high-quality survey that jointly elicits preferences, consumption, income, and wealth in a panel structure.\footnote{Several researchers linked stated preferences to high-quality choice data before \citep{ClausThustrupKreiner_wealyh_beta,Dohmen_AER10_cognitive}. Yet, panel elicitation, as far as we know, only appeared in SHIW and the Preference Parameters Study of Osaka University \citep{Hanaoka_et_al_Japanriskpref} and the Japan Household Panel Survey (JHPS/KHPS) \citep{Hitoshi21}. Although a joint panel of stated preferences and choice data is hard to come by, our identification framework is generalizable to policy stress testing. One can purpose-build surveys eliciting preferences and intended actions in hypothetical scenarios, such as \citep{roth2023effects}.}  We build a spatially granular people-based dataset (in the terminology of \cite{Deryugina2018}) that enables sharp identification of where the earthquake was experienced, based on geospatial seismic intensity data, and where damages occurred, based on official, administratively measured damage records. The survey date of SHIW allows us to identify who was interviewed before and after the earthquake. Such spatially granular data are increasingly used in disaster settings for sharper identification \citep{tellman2021satellite, nakamura2022, halliday2019vog}.

Using difference-in-differences, we first compare those who experienced the earthquake but did not suffer damage, the \textit{shaken but not devastated}, to those who did not experience it. We exclude those whose house was damaged.\footnote{The reason we exclude from our treatment group those who live in damaged municipalities is that their choices between intertemporal monetary payments would be difficult to interpret. Their stated time preferences likely reflect costs of rebuilding and uninsured losses rather than a trade-off between present and future utility. Indeed, in our data, the decline in their stated patience is more than twice that  of the not-devastated group. } We find that experiencing the earthquake made people more impatient: they responded more impatiently to the survey time preference question, and those with sufficient liquidity (non-hand-to-mouth; \citet{Zeldes, kaplan2014wealthy, AguiarBilsCOrina24NBER}) increased nondurable consumption conditional on income by about €175 per month. The increase is substantial: an 11\% rise relative to the pre-earthquake level and to unexposed households, comparable in magnitude to the consumption decline in the first year of unemployment - a benchmark we compute from the same data. These households received no reconstruction transfers, so the increase cannot reflect relief income. 

We design an additional analysis to assess threats to our approach's validity. Our approach rests on an exclusion restriction that the earthquake affects the \textit{shaken but not devastated} only through a preference shock or financial resource shock.  However, households near the epicenter also face indirect exposure through local labor markets, social networks, and equilibrium price changes, channels that are hard to quantify and could confound preference estimates. To partial out these indirect effects, we construct a \textit{spatial control} group: households at distances from the epicenter similar to those of the shaken but not devastated treated group. 
We confirm that the within-distance variation is plausibly as-good-as-random in two ways. First, treated and spatial control households display parallel disposable-income trajectories before and after the earthquake. This means that even if the earthquake impacted them through work, it impacted them similarly. Second, they are also at similar distances from the devastated area, making it plausible that they experienced similar general equilibrium effects on prices.\footnote{We can find treated and spatial control groups at similar distances from the epicenter because seismic waves propagated unevenly due to the area's varied geology. Importantly, many aspects of the propagation were unexpected and only understood ex post \citep{Albano-earthq, Cantore-earthq} - see details in Section \ref{s:earthq}.} Using the richness of our data, we provide further evidence on similarity of some indirect effects between treated and spatial control groups through rental prices, spending on house renovations, and household composition. These estimates also identify a latent impatience shock, indicating that indirect, general equilibrium effects do not drive our results. 

The time preference shock is persistent. Over the three years following the earthquake, the treated group sustains elevated stated impatience and higher consumption, financing the gap by drawing down liquid savings. By year seven, the consumption increase is no longer statistically significant, but liquid wealth is substantially depleted: shaken households, previously at the 57th percentile of the national liquid wealth distribution, fall by 2.6 percentile points.
Consequently, one misfortune leads to the next: households affected by the earthquake end up more financially vulnerable than the control group, despite starting from similar savings buffers.

We assess and find evidence against three alternative mechanisms. First, \cite{AndersenHarrisonLauRutstrom_ECTA08_risk_time} show that stated time preference estimates should also account for risk preferences. In our context, risk preferences could change after the earthquake, which might impact the time preference shock we estimate. We estimate that although risk preferences temporarily increase after the disaster, they revert to baseline within two survey waves. This makes it unlikely that they explain the persistent effects we document. In addition, our results remain the same if we correct the time preference measure with the utility curvature using risk preferences \citep[following][]{Hitoshi21}, or control for risk preferences in the estimation.  Another possibility is that stated time preferences are reflecting beliefs about survival probability - a mechanism documented by \citet{FalkHermleSunde19longevity}. This is a plausible channel, but it cannot explain the magnitude of our findings. Using 2010 ISTAT life tables, we calculate that reconciling the observed decrease in patience through a reduction in survival probability alone would require halving the expected remaining lifespan of a 40-year-old, which is clearly implausible.
 Third, updated beliefs about disaster risk are inconsistent with the absence of any increase in disaster insurance take-up. However, beliefs and preferences are hard to separate. What we observe are intertemporal choices (in hypothetical intertemporal choices and consumption-savings), which beliefs and preferences jointly affect. 
 
 The evidence instead points to a psychological stress mechanism. 
Stress and cortisol exposure have been shown to elevate the salience of near-term outcomes and compress the effective planning horizon \citep{Haushofer2014PsychologyPoverty}.\footnote{Related theoretical work on salience \citep{gennaioli2010comes, bordalo2012salience} shows how attention to salient stimuli distorts decision weights. Originally developed for risky choice, the logic naturally extends to intertemporal decisions \citep{gabaix2021myopia}.}
Consistent with this, stated preference shifts are strongest among groups with the highest stress vulnerability: the less healthy and the elderly. This mechanism is also supported by a decline in self-reported health of the treated group after the earthquake, by a magnitude equivalent to approximately ten years of aging.\footnote{Antidepressant and antipsychotic drug prescriptions also increased after the earthquake, not only among the damaged but even among those who lived further away  \citep{Rossi2011Feb}.} 

These findings identify a previously unmeasured channel of disaster costs: persistent preference disruption from psychological exposure. That such shifts occur in a high-income setting with developed insurance, credit markets, and public safety nets corroborates evidence from developing-country contexts (e.g., \citep{disasters_11floodCambodia_sommarat_2016,disasters_tsunami_thailand_cassar17}), suggesting that preference changes reflect genuine discount-rate shifts, not merely responses to reductions in financial resources.\footnote{Existing research that focused on developed country context did not exclude damaged households \citep{Hanaoka_et_al_Japanriskpref, AKESAKA2019239}. These costs are absent from policy frameworks that focus on direct damages and fiscal relief \citep{NoyJDE09_ndisster_macro, deryugina2017fiscal}.}
As traumatic shocks grow more frequent under climate change, the mechanism we identify may persistently reshape saving behavior, erode wealth buffers, and widen inequality between affected and unaffected households. On a large scale, reduced savings would inhibit intergenerational insurance against climate change. 

The paper is organized as follows. The next section describes our conceptual framework for identifying time preference shocks. Section \ref{s:setting-data} covers the empirical setting and data. Section \ref{s:r-design} covers the empirical design. Section \ref{s:result} provides the estimation results. Section \ref{s:mechanism} discusses potential mechanisms. Section \ref{s:concl} concludes.

 \section{Conceptual Framework}\label{s:framew}
\subsection{Measurement framework}
To fix ideas, suppose we have a panel of stated time preferences, elicited through hypothetical ``money earlier or later'' (MEL) choices, where subject $i$  at time $t$ chooses between smaller immediate and larger delayed rewards. MEL choices, expressed as discount factors $\tilde{\beta}_{i,t}$, are biased measures of the latent $\beta_{i,t}$: low income or financial resources cause lower stated patience in MEL answers \citep{ambrus2015, Epper2015}, even when financial resources are low only temporarily \citep{carvalho2016poverty}.\footnote{\cite{guiso2008trusting-stockmkt} was among the first to highlight how financial constraints shape responses to hypothetical questions in the context of risk perception and portfolio choice. Other contexts have also shown common biases in hypothetical choices, from buying intentions \cite{manski1990intentiontobuy} to fertility perceptions \cite{Paulavalente2024fertility}. For a collection of papers on biases in hypothetical choices, see \cite{Almas2024} Appendix A1.} Yet MEL choices still carry valuable information about the true underlying time preference: those with lower stated patience save less \citep{FalkQJE18},  accumulate less wealth \citep{ClausThustrupKreiner_wealyh_beta}, and invest less in human capital \citep{Sunde2022Patience}.
The following measurement equation summarizes these:
\begin{align}\label{eq:beta}
    \tilde{\beta}_{i,t} = B(\underset{ +}{\vphantom{\beta}Y_{i,t}},X_{i,t}; \underset{+}{\beta_{i,t}}),
\end{align}
where the signs below the arguments of $B(\cdot)$ indicate the sign of the corresponding partial derivatives.  $\tilde{\beta}$ increases both in disposable income $Y$ and the true underlying discount factor $\beta$.
$\frac{\partial B}{\partial Y}>0$ is testable, and we verify that it holds in our data in Section~\ref{subsec:bias} by testing the implied covariance structure (deriving covariances similar to \cite{Blundell2008}). $X$  collects further correlates, such as education, age, and family size, for which we do not specify the sign of the partial derivatives.%

Households choose consumption according to a standard intertemporal consumption-savings model: maximize lifetime utility $\mathbb{E}_0 \sum_{t=0}^{\infty} \beta^t u(C_{i,t})$, where $u(\cdot)$ is strictly increasing, strictly concave, and satisfies standard Inada conditions. Households allocate resources across periods using a risk-free savings technology with return $(1+r)$, subject to the intertemporal budget constraint $A_{i,t+1}=(1+r)A_{i,t}+Y_{t}-C_{i,t}$, where $ A_{t}, Y_{t}, C_{t}$ denote liquid wealth, disposable income, and consumption at time $t$, respectively. The properties of the optimal consumption policy function are well established: 
\begin{align}\label{eq:c}
    C_{i,t}=C(\underset{0+}{\vphantom{\beta}Y_{i,t}},X_{i,t}; \underset{-}{\beta_{i,t}}).
\end{align}
A decline in income weakly reduces consumption, denoted by $0+$.\footnote{Theoretically, households raise consumption in response to permanent positive income shocks, but not in response to transitory shocks (\cite{Hall1978, Carroll1992, CarollHolmKimball21_c_liq, AguiarBilsCOrina24NBER, Commault2022}). Empirically, households do raise consumption after transitory income shocks (see, for example, \cite{fagereng2021mpc, Commault2022}). Similarly, households theoretically should not respond to anticipated income changes, but \citep{hsieh2003consumers}, for example, finds that they do.} Consumption increases when $\beta$ is lower, as future utils are less valued. The discount factor $\beta_{i,t} \in (0,1)$ may vary over time; we implicitly assumed these variations are unexpected by the individual, i.e. $\beta_{i,t}=\mathbb{E}_t \beta_{i,t+k}, k>0$.

\begin{theorem}
    A latent decline in the discount factor $\beta_{i,t}$ can be distinguished from a decline in disposable income by the co-movement of stated preferences and consumption, assuming an unconstrained consumption choice:
    \begin{itemize}
        \item[i] An impatience time preference shock (a decline in $\beta_{i,t}$): 
$\Delta \tilde{\beta}_{i,t}<0$ and $\Delta C_{i,t}>0$.
\item[ii] A liquidity shock (a decline in disposable income $Y_{i,t}$): $\Delta \tilde{\beta}_{i,t}<0$ and $\Delta C_{i,t}\leq 0$.
    \end{itemize}
   
Hence, identification of time preference shock relies on the sign restriction
$sign(\Delta \tilde{\beta}_{i,t}) = - sign(\Delta C_{i,t})$.
    \end{theorem}

    Our identification requires households to be on their Euler equation, and we therefore exclude hand-to-mouth (HTM) households, who cannot raise consumption in response to a decline in $\beta_{i,t}$. Symmetrically, a genuine increase in patience is identified when stated patience rises and consumption falls. In this case, HTM might not reduce consumption, given that their consumption is already below the unconstrained optimum. 

\begin{figure}[h!] 
    \centering
   \begin{minipage}{0.49\textwidth}
        \resizebox{.9\linewidth}{!}{\begin{tikzpicture}

    \definecolor{mapgreen}{RGB}{152,185,100}

    \draw[thick, ->] (0,0) -- (7,0) node[right] {$Y$};
    \draw[thick, ->] (0,0) -- (0,4.5) node[above] {$C_0$};

    \draw[thick] (0,0) -- (6,3);

    \draw[line width=2pt, dashed, mapgreen] (0,0) -- (6,4);

    \fill (3,1.6) -- (3.15,1.8) -- (3.3,1.6) -- cycle;
    \fill (3.05,1.6) -- (3.05,1.4) -- (3.25,1.4) -- (3.25,1.6) -- cycle;
    
    \fill (5,2.6) -- (5.15,2.8) -- (5.3,2.6) -- cycle;
    \fill (5.05,2.6) -- (5.05,2.4) -- (5.25,2.4) -- (5.25,2.6) -- cycle;

    \fill (5,3.5) -- (5.15,3.7) -- (5.3,3.5) -- cycle;
    \fill (5.05,3.5) -- (5.05,3.3) -- (5.25,3.3) -- (5.25,3.5) -- cycle;

    \node at (3.3,1) {\textbf{$Y$ decline}};
    \node at (5.2,4.1) {\textbf{$\beta$ decline}};

    \draw[dashed] (3.15,0) -- (3.15,1.8);
    \node at (3.15,-0.2) {$Y'$};

    \draw[dashed] (5.15,0) -- (5.15,3.7);
    \node at (5.15,-0.2) {$Y$};
    
    \draw (3.15,0.1) -- (3.15,-0.1);
    \draw (5.15,0.1) -- (5.15,-0.1);

    \draw (0,0) -- (2,1);
    \draw (2,1) -- (4,2);
    \draw (4,2) -- (6,3);

    \node[right] at (6.1,3) {\large $C_0=\frac{1}{1+\beta}Y$};

    \node[right] at (6.1,4) {\large $C_0=\frac{1}{1+\textcolor{mapgreen}{\mathbf{\beta'}}}Y$};
\end{tikzpicture}}
   \end{minipage}
        \caption{Consumption policy function for log utility and deterministic income, as a function of discounted lifetime income.}
    \label{fig:logutil}
\end{figure}
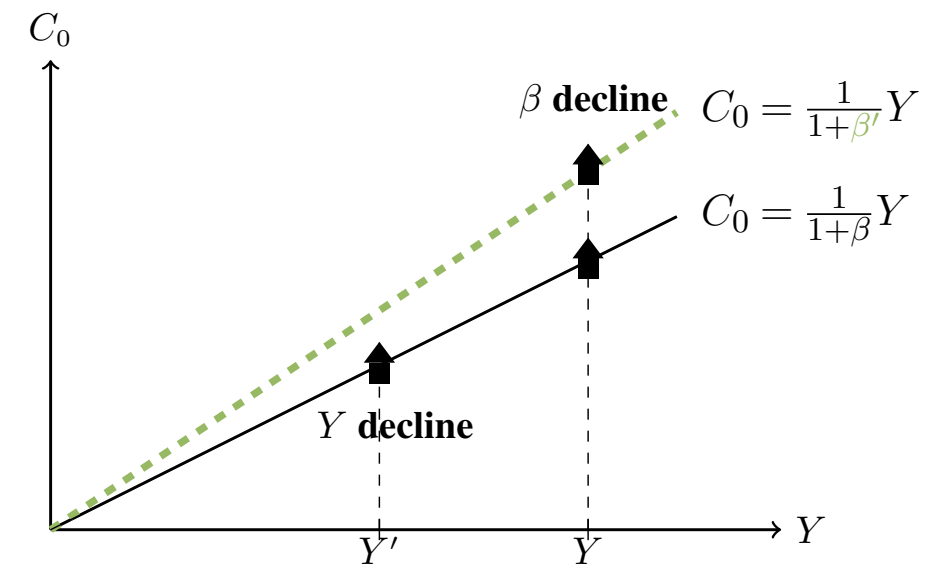

As an illustration, we plot a simple log-utility case in Figure \ref{fig:logutil},  with deterministic income, for which a closed-form consumption policy function exists: $ C_{i,0}= \mathbb{E}_0\frac{1}{1+\beta}  \sum_{t=0}^{\infty}\frac{Y_{i,t}}{(1+r)^t}=\frac{1}{1+\beta} Y$,  with $Y$ denoting discounted lifetime income.
 A decline from $Y$ to $Y'$  lowers the stated discount factor via the bias in stated preferences, and shifts consumption downward along the existing policy function.\footnote{If both income and the latent time preference decline, consumption would still increase conditional on income.} In contrast, a decline in $\beta$ reduces the stated discount factor, but pivots the policy function upward, increasing immediate consumption, assuming an unconstrained choice. We identify time preference shocks by comparing two observationally similar individuals pre-event: if the treated individual subsequently states lower patience after the event and increases consumption, this jointly identifies a decline in the true discount factor. 

\subsection{Application to earthquakes}
Natural disasters pose two identification challenges. First, destruction may push households off their Euler equation. Second, the disaster affects disposable income directly, so conditioning on income would induce
bad-control bias \citep{angrist2014mastering}, as income would absorb part
of the treatment effect.

We address the first challenge by excluding households with damaged homes from the treatment group, using geocoded seismic and damage data (red and
yellow in Figure~\ref{fig:damage}). Our \textit{treated} group is \textit{shaken but not devastated}; our \textit{control} group consists of households that experienced no ground motion.

We address the second challenge using the disaster's spatial structure. Disaster effects on income spread through labor markets, networks, and
general equilibrium channels well beyond the affected area
\citep{barrot2016, carvalho2020}. 
We therefore choose a control group, we term \textit{spatial control} that share
treated households' pre-disaster characteristics and spillover exposure, so common income shocks difference out.

\begin{theorem2}
 The spatial control group $\mathbb{S}_{C_s}$ is constructed by households, that compared to the shaken but not devastated treatment $\mathbb{S}_T$ are
 \begin{itemize}
     \item[i] similar in observables before the earthquake (Spatial balance assumption)
      \item[ii]  have parallel trends in disposable income $\mathbb{E}\!\left[\Delta Y_{it} \mid i \in  \mathbb{S}_{T}\right]
    = \mathbb{E}\!\left[\Delta Y_{it} \mid i \in \mathbb{S}_{C_s}\right]
     \quad \forall t$,
     \item[iii] at similar distances from the damaged area.
  \end{itemize}
 \end{theorem2}
  We operationalize this by restricting the control group  to  reside within the same distance $d(r_i, s)$ from the epicenter of the earthquake $s$ as the treated:
 \begin{align}
     \mathbb{S}_{C_s} = \{r_i \in C : d(r_i, s) \leq d^*\},
     \quad \text{where } d^* = \max_{r_j \in T} d(r_j, s),
 \end{align}
where $r_i \in \mathbb{R}^2$ denotes location (latitude, longitude) of household $i$, $s\in \mathbb{R}^2$ is the epicenter of the earthquake. We verify that the \textit{spatial control} group satisfies \textit{i.} in Section \ref{sec:est}, \textit{ii.} in Section~\ref{subsec:results-spatial}, and \textit{iii.} in Section \ref{subsec:results-spatial}. Condition \textit{i.} ensures that treatment and control have similar characteristics before the earthquake. Condition \textit{ii.} ensures that disposable income, which is causing a bias in stated preferences, evolves similarly both before and after the earthquake. Relative to the spatial control group, income is therefore not affected by treatment, and conditioning on it does not absorb part of the treatment effect: income is not a bad control in this comparison. The logic of condition \textit{iii} is that prices typically increase in affected areas \citep{MariassuntaGiannetti2026who}, but when treated and control are at a similar distance from the damaged area, they likely face similar price increases. We examine rental prices, and also the role of closeness to damaged areas in Section \ref{sec:cons}.

Conditions \textit{i} to \textit{iii} jointly amount to a relaxation of the standard Stable Unit Treatment Value Assumption (SUTVA) by explicitly accommodating spatial interference. Rather than assuming no interference, we absorb spatial spillovers into the control group. Our identification logic is similar to the spatial inference approach of \citet{bernheimBjorkegrenNaeckerPollman2021causal}, which leverages proximity to mitigate local interference.

\subsection{Advantages and Limitations}

 Our identification framework has both methodological advantages and practical limitations relative to existing approaches.\\

\noindent \textit{Minimal structural assumptions} A key advantage is that identification relies only on sign restrictions derived from standard consumption theory, which requires no parametric assumptions on the utility function.
By contrast, recent work that made progress combining stated and revealed preferences for causal inference requires stable mappings between these two \citep{bernheimBjorkegrenNaeckerPollman2021causal, meango2025combiningstatedrevealedpreferences, Almas2024}, as does structural estimation of 
preference parameters \citep*{Gourinchas2002}. \\

\noindent \textit{Data requirements} This generality comes at a cost: the framework requires joint observation of stated and revealed preferences in a panel structure, a combination rarely available outside specialized surveys like the SHIW. That said, the requirement is less restrictive than it may appear. Purpose-built evaluations can elicit both hypothetical choice questions and intended behavioral responses within a single survey. Such an approach was taken, for example, in  \citet{roth2023effects}, who present respondents with hypothetical monetary policy scenarios and measure planned consumption responses. Such designs would extend the framework to a wide range of policy-relevant shocks. \\

\noindent \textit{Sample restrictions and scope} The main limitation of our approach is that it relies on households being on their Euler equation,  that is, able to adjust consumption in response to a time-preference shock. This restriction,
however, sharpens identification. As \citet{carvalho2016poverty} show, liquidity constraints can confound
measured time preferences. By restricting attention to unconstrained households, we separate the two. 

Our sign restrictions identify the direction of preference changes, but not the magnitude. To assess the magnitude, additional functional form assumptions are needed. 

 \section{Empirical Setting and Data}\label{s:setting-data}

\subsection{The L’Aquila Earthquake}\label{s:earthq}
In the early morning of April 6, 2009, at 3:32 a.m., a medium-power earthquake ($M_w=6.3$) struck the city of L’Aquila and nearby municipalities in central Italy. The main shock occurred just four kilometers southwest of L’Aquila at a shallow depth (8-9 km), causing severe structural damage. Occurring at night, it caught many residents in bed; 309 lives were lost, 1,500 were injured and 70,000 displaced.\footnote{Survivors cited a false sense of security from a government press conference days earlier, where officials downplayed the likelihood of a major shock. The Department of Civil Protection convened a seismologist panel largely for public reassurance; statements made contradicted the committee’s actual conclusions.} In the aftermath, the government evacuated affected populations and resettled them in camps and hotels, but by winter, they were placed into government-constructed new anti-seismic housing next to L’Aquila.\footnote{Families could instead choose a lump-sum payment, but most opted for the newly-built houses. The construction was financed using European Union funds allocated for temporary accommodations, but many still lived in these houses during our visit in 2024 \citep{EU2013}.} There was some damage to roads, infrastructure, but the repairs were swift. The reconstruction of L'Aquila, on the other hand, was slow. L’Aquila, once a vibrant gathering place, remained closed for years, undermining social life and economic activity.

\begin{figure}[h!] \centering
	\includegraphics[width=0.95\textwidth]{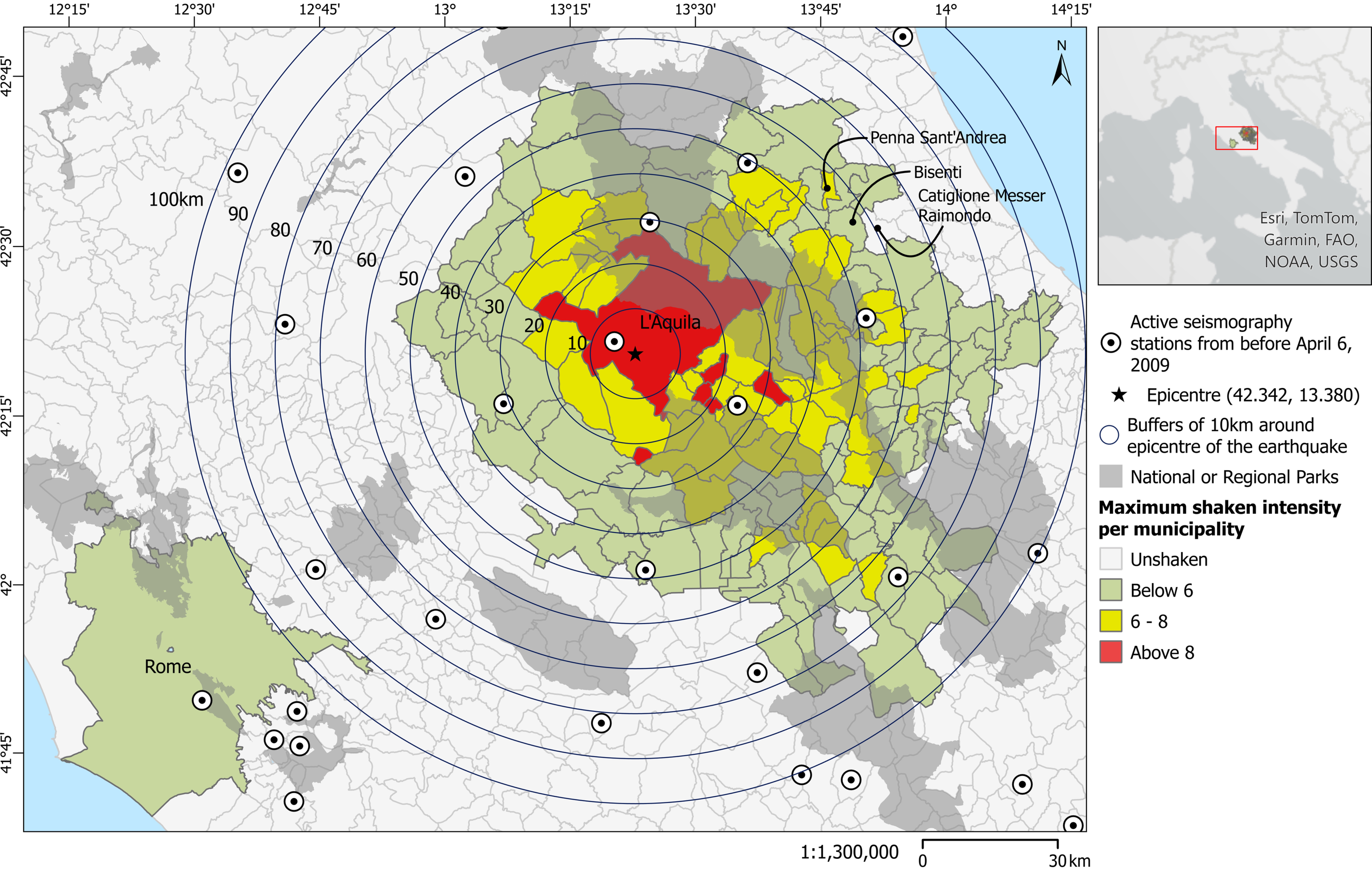}
	\caption{Seismic Activity from L'Aquila Earthquake.\label{fig:damage}}
    \legend{Data: INGV.  Municipalities shaded by Modified Mercalli Intensity:  >8 severely damaged, 6-8 damaged, below 6 shaken but not damaged. Unshaded municipalities recorded no shaking.}
\end{figure}

The timing of earthquakes is impossible to forecast.\footnote{Italy's seismicity stems from the slow collision and rotation of the African and Eurasian tectonic plates. Faults slip when geological stress surpasses local thresholds, but predicting where and when this occurs is not possible.} Hazard maps summarize the riskiness of areas, but in the case of L’Aquila, it was an unexpected epicenter because it was not categorized as high-risk prior to the event, ranking only as “moderate” (level 2 on a four-point national hazard scale). In addition, the region had not experienced an earthquake of similar magnitude in nearly a century, and the shock activated faults absent from
existing hazard maps.\footnote{Hazard maps were revised afterward, and the
event prompted new investment in seismic monitoring under Law Decree
no.~39/2009.} 

Shaking intensity also varied sharply at equal distances from the epicenter (Figure~\ref{fig:damage}).  For instance, Penna Sant’Andrea and Bisenti lie roughly 40 km from the epicenter, yet the former suffered damage (MMI 6), while the latter did not (MMI 5). A nearby municipality, Castiglione Messer Raimondo, recorded no measurable shaking. Our treated group, shaken but not devastated (shaded green), lies along fault lines and in areas with looser soil. Therefore, equidistant from the epicenter,  we find treated and non-treated municipalities as well.\footnote{The shallow depth,  amplified surface-level impacts, and differences in soil composition generated differences in the propagation of seismic waves \citep{Albano-earthq, Cantore-earthq}.  This pattern, was understood only ex-post, and was incorporated into hazard maps later. Extremely damaged areas with MMI scale between 8 and 9.5 experienced widespread structural failure, including collapsed buildings and fatalities (shaded red). Locations with moderate damage (e.g., broken windows, chimney collapse, failure of non-reinforced structures) were assigned MMI values between 6 and 8 (shaded yellow). Our treatment group, the shaken but not devastated is shaded green. They had perceptible shaking without structural damage, with a Modified Mercalli Intensity (MMI) between 4 and 5.5.} Because propagation depends partly on soil composition, which may correlate with urbanization, we show that treated and control municipalities are balanced on pre-earthquake characteristics (Table~\ref{tab:stat}). Because Rome, the metropolis, is in our treatment group alongside smaller Abruzzo communities, we examine the role of Rome in our robustness section \ref{subsec:results-rob}. The earthquake thus generated
two sources of as-good-as-random variation, that form the basis of our identification: the timing and location of the
epicenter, and the heterogeneous propagation of seismic energy.

Aid flowed into the region in the aftermath of the earthquake, which could confound the estimates if it was an income transfer to the treated group. Reconstruction funds, subsidies, and tax suspensions were, however, restricted to municipalities with
MMI 6 or above (Law Decree no. 39/2009). Our treated group received none.\footnote{The government financed the restoration of L’Aquila, but reconstruction was slow (remains incomplete as of 2026) and riddled with allegations of corruption and mafia infiltration. The government paid the construction companies carrying out the work directly, rather than financing households \citep[see][]{acconciaCorsetti2020liquidity}.} 

 \subsection{Seismic and Damage Data} 

Our earthquake data source is the \textit{Catalogo Parametrico dei Terremoti Italiani},\footnote{Publicly available via the Istituto Nazionale di Geofisica e Vulcanologia (INGV).} Italy’s dense monitoring network gives information about seismic movements, and the damage data was collected by the \textit{Quick Earthquake Survey Team} (QuEST). We classify a municipality as \emph{shaken but not devastated treatment} if its maximum MMI lies between 4 and 5.5, and as damaged if it is MMI 6 and above (Figure \ref{fig:damage}). 

To ensure the reliability of our data, we contacted QuEST to understand their procedures. Importantly, the location and extent of seismic movement were not inferred ex post: they were observed in real time through seismographs (dots on Figure \ref{fig:damage}), which provided immediate and spatially precise data on ground motion. This allowed QuEST to prioritize affected areas, cataloguing damage severity, building vulnerability, and structural impacts, which were summarized into MMI scores.%
 We cross-validated and confirmed that no seismically active location is missing from the damage dataset. 
 The combination of objective instrumentation and comprehensive fieldwork makes these data highly reliable.

\subsection{Household Data} \label{subsec:shiw}
Our primary data source on households is the Survey on Household Income and Wealth (SHIW), of Banca d'Italia.
SHIW has been a key resource in life-cycle research due to its high quality and distinguishing feature: it surveys income, consumption, and savings within a single survey  \citep[see, for example,][]{Guiso1996,attanasio2001intertemporalSHIW,Jappelli2010,jappelli2000dynamics}.\footnote{The high data quality is ensured by trained enumerators. It is conducted through in-person interviews and by allowing households to verify their answers via tax records and bank statements. The survey’s representativeness further enhances its quality, achieved through stratified random sampling and replacing attrited households with similar ones.} The SHIW surveys about 8,000 households biennially, with $50\%$ in a rotating panel.\footnote{The SHIW follows a rotating panel 
design in which approximately 50 percent of households are re-interviewed 
across waves, with panel households selected randomly from those 
interviewed in the previous edition \citep{BankofItaly2022}. The Bank 
of Italy explicitly adjusts survey weights to account for panel attrition 
by stratifying the panel component ex-post by income class and 
head-of-household employment status. Since we apply these survey weights 
throughout our analysis, general panel attrition is unlikely to bias 
our estimates.} 

Our data definitions follow the life-cycle literature \citep[see][]{Attanasio1995}; savings is residual disposable income net of consumption (durable and non-durable). Net liquid assets are deposits, bonds, and equity minus outstanding debt. Cash on hand is disposable income plus net liquid assets. See detailed data definitions in Appendix \ref{appx:data}.\\

\noindent \textit{Time Preferences} 
To identify time preference shocks, we focus on a previously untapped section of SHIW that elicits time preferences via a hypothetical intertemporal trade-off question (``money earlier or later '', see Section \ref{appx:df}). This module was administered in four waves: two waves before the earthquake (2004 and 2008) and two after (2010 and 2012). The panel structure of the data is especially valuable, as it allows us to track stated preferences over time and observe their evolution relative to the earthquake.
\begin{figure}[h!]
\centering
\subfigure[Stated Discount Factor]{\label{fig:a}
\includegraphics[scale=0.3]{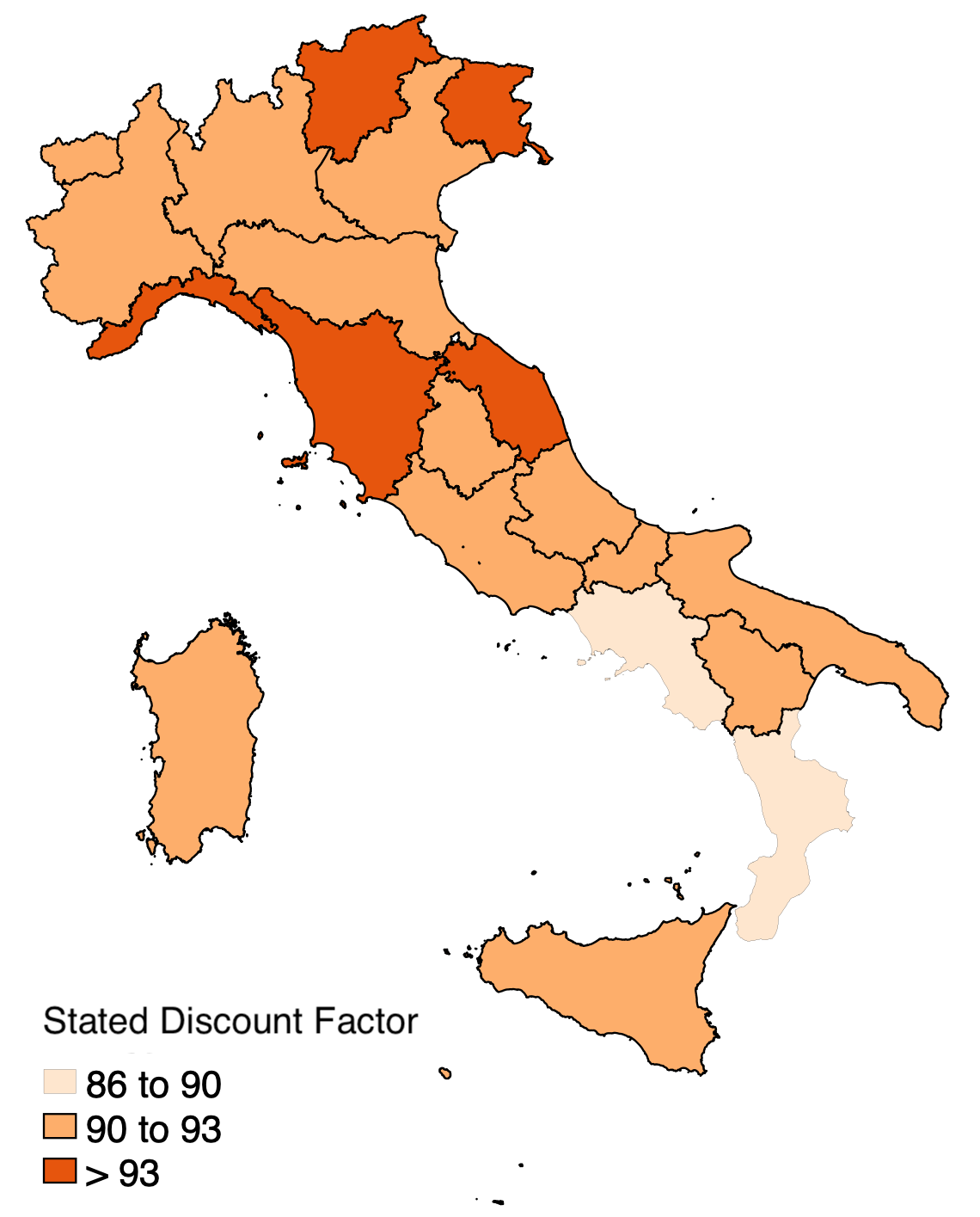}}
\subfigure[Income]{\label{fig:b}
\includegraphics[scale=0.3]{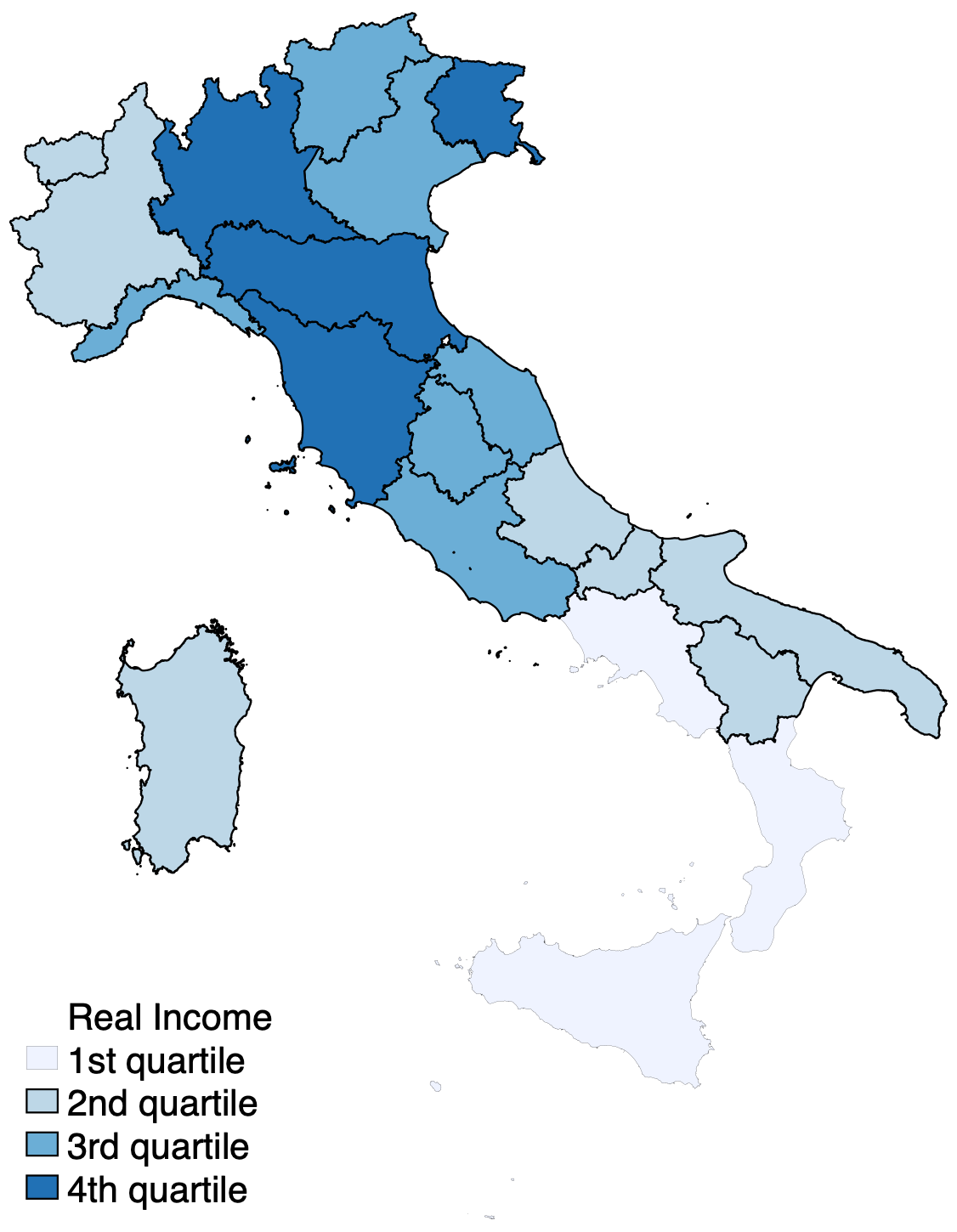}}
\caption{Stated Discount Factor and Income Quartile Across Italian Regions (Pre-Earthquake) \label{fig:map_pref}}
\legend{Panel (a) colors the 20 Italian regions by their pre-2009 average stated time preference, and panel (b) by their average position in the real income distribution. Darker colors represent a higher stated discount factor (more patient) and higher income.} 
\end{figure}
Although various methods exist for eliciting time preferences, the SHIW question is both simple and experimentally validated, achieving a response rate of 88 percent. One exception is the 2008 wave, when time preference was elicited only for half of the households, for heads born in even-numbered years.\footnote{As a data quality check, Figure~\ref{fig:att} 
confirms that response frequency to the patience question is uncorrelated 
with patience and education, suggesting no systematic pattern of 
non-response in the raw data.} We express MEL responses as implied discount factors, where a decline reflects reduced patience (see Appendix \ref{appendix:pref}). 

Figure \ref{fig:map_pref} illustrates the countrywide data. Stated time preferences reflect more patience in the north, where average household income is also higher, compared to the South.\footnote{The same pattern holds across households: those with higher income are more patient; see Figure \ref{fig:df_binsreg}.} 

Time preference data are collected from only one household member, typically the household head, raising the question of whose time preference governs household intertemporal choices \citep{Almas2022handbook}. In our baseline estimates, we treat all answers as characteristics of the household. For robustness, we restrict the sample to households with the same respondent across waves  (Section \ref{subsec:results-rob}).  

\subsection{Identifying earthquake exposure}\label{subsec:exposure}
We link SHIW households to seismic exposure at the municipality level. Treatment is assigned 
by residence prior to the earthquake, to mitigate concerns about endogenous relocation. For households first observed in 2010 or later, we use self-reported duration of residence to determine pre-earthquake location. Because SHIW interviews are conducted in person, self-reported residence provides a credible measure for actual residence.%

\subsection{Timing and estimation sample}
\begin{figure}[h!]
    \centering
\begin{tikzpicture}\small
\centering
			\draw  (-5,0) -- (5,0);
			\filldraw [gray] (-2,0) circle (2pt) (-2,0.5) node {2006} ;
			\filldraw [blue]   (-4,0) circle (3pt) (0,0) circle (3pt) (2,0) circle (3pt)  (4,0) circle (3pt)
			(4,0.5)  node  {2012}
			(2,0.5)  node  {2010}
			(0,0.5)  node  {2008}
			(-4,0.5)  node  {2004}  ;
			\draw [red] (1.27,1) node[above] {6 April 2009. L'Aquila earthquake} -- (1.28,0)  ;	
						\filldraw [black] (1,0) circle (2pt) (1,-0.5) node {2009 } (1,-0.8) node {\textbf{Interviews} } ;
			\draw [->] (1,0) arc [start angle=0, end angle=-150, radius=10pt];
		\filldraw [black] (-3,0) circle (2pt) (-3,-0.5) node {2005};
		\draw [->] (-3,0) arc [start angle=0, end angle=-150, radius=10pt];
			\filldraw [black] (-1,0) circle (2pt) (-1,-0.5) node {2007};
		\draw [->] (-1,0) arc [start angle=0, end angle=-150, radius=10pt];
			\filldraw [black] (3,0) circle (2pt) (3,-0.5) node {2011};
		\draw [->] (3,0) arc [start angle=0, end angle=-150, radius=10pt];
		\filldraw [black] (5,0) circle (2pt) (5,-0.5) node {2013};
		\draw [->] (5,0) arc [start angle=0, end angle=-150, radius=10pt];
	\end{tikzpicture}
        \caption{Timing of SHIW Surveys. \label{fig:timing}}
        \legend{Interviews are conducted in years shown in black, surveying previous year economic outcomes, i.e. interviews for the 2004 wave are conducted in 2005. Time preference questions were asked only in waves colored blue (2004, 2008, 2010, 2012).}
    \end{figure}
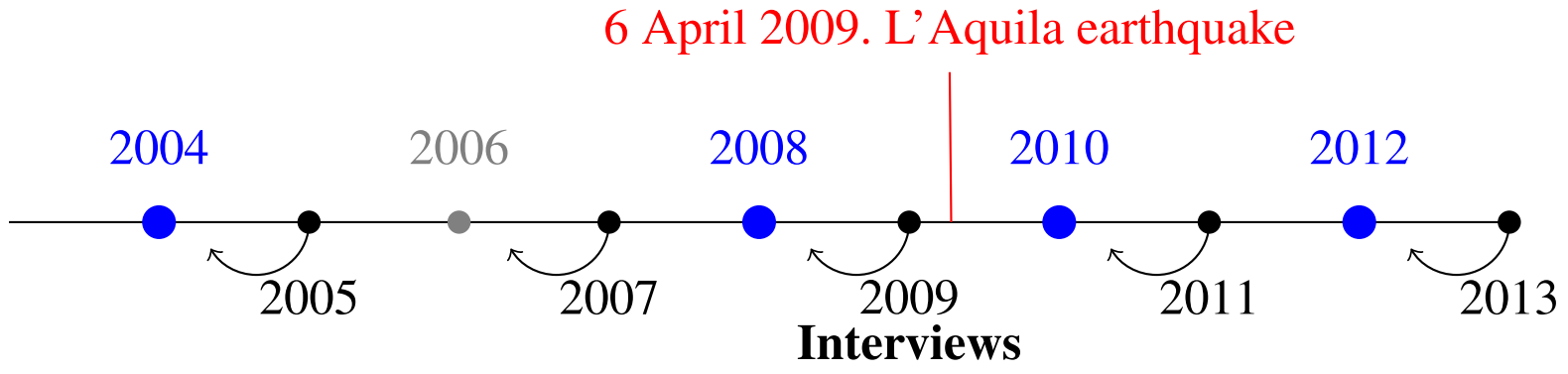

Income and consumption refer to the previous calendar year, so in the 2008
wave (fielded in 2009) they are pre-earthquake outcomes
(Figure~\ref{fig:timing}). Stated preferences, by contrast, are elicited at
the interview, and because fieldwork straddled April 6, 2009, interview
dates let us measure them immediately before and after the earthquake.

As a baseline, we examine stated time preferences on a panel of household heads aged 25 to 90 at the time of the earthquake (5,029 households); as a robustness check, we repeat the analysis on a subsample of working-age household heads,  25–70 (3,736 households).  
For consumption–savings estimations, we restrict the sample to working-age heads, following standard life-cycle exclusions \citep{Deaton1991,Attanasio1995}, and expand the survey years to 2002-2016, yielding an unbalanced panel of 4,645 households. 

\subsection{Empirical Validation of the bias in stated preferences}\label{subsec:bias}

Our identification rests on $\frac{\partial B}{\partial Y}>0$ in equation~\eqref{eq:beta}. We now provide direct evidence for this condition using pre-earthquake panel variation. The cross-sectional correlation between income and stated time preference documented, e.g., in \citep{Haushofer2014PsychologyPoverty,carvalho2016poverty,guiso2008trusting-stockmkt, ClausThustrupKreiner_wealyh_beta} is not sufficient, because this data pattern is also consistent with stable individual heterogeneity rather than a bias in stated time preferences. Identification requires the stronger condition: stated preferences respond to income changes within individuals over time. 

We follow the method of \cite{Blundell2008}, and derive the covariance structure implied by a model with income as a random walk plus a transitory shock, and a linear version of the measurement equation \eqref{eq:beta}.
\begin{align}
\begin{aligned}
    \tilde{\beta}_{i,t} &= \alpha_0+\alpha  y_{i,t} +\tilde{\epsilon}_{i,t}\\
     y_{i,t}&=\mu_{i,t}+\varepsilon_{i,t}\\
    \mu_{i,t}&=\mu_{i,t-1}+\nu_{i,t},
\end{aligned}\label{eq:beta2}
\end{align}
where $y_{i,t}$ is log-income, composed of a random walk permanent component $\mu$ and an i.i.d. transitory component $\varepsilon$. The shock innovations  $\nu_{i,t}$ and  $\varepsilon_{i,t}$ are independent, have zero expected value and variances $\sigma^2_\nu$ and $\sigma^2_\varepsilon$.\footnote{For simplicity, in equations \ref{eq:beta2} we abstract from life-cycle characteristics  $X_{i,t}$. In the empirical specifications controls include household head age, education, and liquidity proxies.}

Under these assumptions, the model yields the following set of testable predictions:
\begin{equation}
    \begin{aligned}
     Cov (\tilde{\beta}_{i,t},y_{i,t})&=\alpha Var (y) \\
 Cov (\tilde{\beta}_{i,t},y_{i,t+1})&=\alpha Var(\mu)< Cov (\tilde{\beta}_{i,t},y_{i,t})\\
 Cov (\tilde{\beta}_{i,t},\Delta y_{i,t+1})&=-\alpha\sigma^2_\varepsilon\\
 Cov (\tilde{\beta}_{i,t},\Delta y_{i,t+2})&=0.
 \end{aligned}\label{eq:cov}
 \end{equation}
Table \ref{tab:cov} confirms all four moment conditions: the discount factor correlates more strongly with current than future income (columns 1–2), is negatively associated with next-period income growth (column 3), and statistically does not correlate with two-period-ahead growth (column 4), consistent with attenuation bias from transitory shocks.
\begin{table}[h!]
 \caption{Covariance of stated time preference with income}
\label{tab:cov}
\begin{tabular}{@{}lcrcrrr@{}@{}}
\hline
 &\multicolumn{2}{c}{Income Level}& ~ &\multicolumn{2}{c}{Income Change}          \\
  &\multicolumn{1}{c}{(1)} &\multicolumn{1}{c}{(2)} &&\multicolumn{1}{c}{(3)} &\multicolumn{1}{c}{(4)}          \\
 \cline{2-3}\cline{5-6}
 &\multicolumn{1}{c}{$y_t$}&\multicolumn{1}{c}{$y_{t+1}$}& ~&\multicolumn{1}{c}{$\Delta y_{t+1}$}&\multicolumn{1}{c}{$\Delta y_{t+2}$}\\
&&&&\\
$\tilde{\beta}_{i,t}$            &       2.405&       2.127 & ~  &      -0.673&      -0.388         \\
            &     (0.124)         &     (0.150)   & ~      &     (0.192)         &     (0.282)         \\
&&&&\\
Observations       &       27,699         &       15,142    & ~     &       14,978         &       10,651         \\
\bottomrule
\end{tabular}
\legend{Estimated covariance of stated discount factors with income levels (columns 1-2) and income changes (columns 3-4).}%
\end{table}

In addition, in Online Appendix Figure~\ref{fig:event_cons_preE} we show that household consumption behavior is consistent with the buffer-stock saving model of \cite{carroll1996concavity}. Non-durable consumption rises with income but less than proportionally, yielding a declining consumption-to-income ratio, and those with more impatient stated time preferences consume a larger share of their income.

\section{Research Design}\label{s:r-design}
Our goal is to identify preference shocks by exploiting regional variation in exposure to the L’Aquila earthquake.  In this section, we first show the importance of assigning treatment based on whether people experienced seismic movements but did not suffer damage. Then we explain our DiD and event study design.

\subsection{Treatment Assignment: Proximity versus Experience }\label{subsec:spatialtreatment}
We start by comparing our treatment assignment to a common approach in the literature, which proxies disaster experience by geographic proximity to the epicenter.  
We estimate the following regression
on post-earthquake data: 
\begin{equation}
\tilde{\beta}_{imt} = \alpha_{1}\indicator_{m \in \textit{Earthquake}} + \mathbf{X}^{'}_{it} \alpha_X +\indicator_t
+\epsilon_{imt}, \label{reg:nakamura}
\end{equation}

\noindent where $\tilde{\beta}_{imt}$ denotes the stated discount factor for household head $i$, residing in municipality $m$, and interviewed in year $t$. The indicator $\indicator_{m \in \textit{Earthquake}}$ captures alternative exposure definitions: seismic exposure or proximity to the epicenter. $\mathbf{X}^{'}_{it}$ includes household and individual covariates (e.g., age, gender, education, household size, income, and credit constraints). Year fixed effects $\indicator_t$ absorb time-specific aggregate shocks. Standard errors are clustered at the municipality level, and all regressions use SHIW survey weights. We pool panel and cross-sectional data, using post-disaster cross-sectional data. Observing only post-disaster data is common, and this specification also maximizes statistical power. 

\begin{table}[hbt!]
\begin{center}
\caption{Effect of Earthquake on Stated Discount Factor - OLS (Post-Earthquake Data)}
 \label{tab:nakamura}
\begin{tabular}{l*{4}{c}}
\hline
                    &\multicolumn{1}{c}{(1)}&\multicolumn{1}{c}{(2)}&\multicolumn{1}{c}{(3)}&\multicolumn{1}{c}{(4)}\\
 \hline \small
0 to 40 km             &      -3.891&      -5.182&                     &                     \\
                    &     (1.268)         &     (0.726)         &                     &                     \\
40 to 125 km            &       0.122         &       0.615         &       0.122         &                     \\
                    &     (0.490)         &     (0.444)         &     (0.489)         &                     \\
\{40 to 125 km\} X Shaken      &                     &      -1.665&                     &                     \\
                    &                     &     (0.385)         &                     &                     \\
Shaken  (excl. $\leq$ 40 km)       &                     &                     &                     &      -1.064\\
                    &                     &                     &                     &     (0.233)         \\

&&&&\\
Mean of dep. var.       &        91.6         &        91.7         &        91.6         &        91.7         \\
Observations        &       15,973         &       15,526         &       15,842         &       15,413         \\
$R$-squared           &       0.036         &       0.036         &       0.036         &       0.035         \\
\hline
\end{tabular}
\legend{OLS estimates of the earthquake’s impact on stated discount factors (2010 and 2012 data). Sample: household heads between 25 and 90. Treatment: distance to epicenter. Controls: age, education, income, credit constraint, wealth, and family size, year fixed effects. Standard errors are clustered at the municipality level; survey weights applied.}
  \end{center}
\end{table}

Column 1 of Table \ref{tab:nakamura} reports results based on distance to the epicenter, dividing municipalities into an \textit{inner ring} (within 40 km, with substantial destruction), an \textit{outer ring} (40–125 km, variable seismic activity), and a control group (more than 125 km, no ground motion).
Households in the inner ring report discount factors roughly 3.9 percentage points lower than those further away, a decline of about 4\% relative to the pre-earthquake mean. This finding is consistent with prior evidence that links disaster proximity to changes in preferences \citep[e.g.,][]{disasters_tsunami_thailand_cassar17, KUROISHI2024}.

The outer ring, however, seems to mask substantial heterogeneity. The average effect is null (columns 1, 3), yet shaken municipalities within the outer ring  report discount factors 1.7 points lower than unshaken ones (column 2). Distance to the epicenter alone therefore fails to capture the treatment effect.

A second observation is that those who experienced major damages in the inner ring stated much lower discount factors compared to the shaken but not devastated group, 3.9 points versus 1.7 points lower than the unshaken group. This gap suggests the inner-ring effect largely reflects financial hardship due to the devastation rather than the psychological experience itself. This motivates our experience-based \textit{shaken but not devastated} treatment definition (column 4), excluding damaged households, where destruction likely caused a downward bias on stated time preferences. 
\subsection{Estimation Strategy}\label{sec:est}
We implement our identification framework outlined in Section~\ref{s:framew} in two steps:
(i) a difference-in-differences (DiD) model exploiting within-household changes over time, and
(ii) an event study that traces the dynamic evolution of preference shocks and associated consumption-savings behavior. 
Each approach offers complementary strengths:%
DiD controls for time-invariant household heterogeneity, and the event study tests timing and persistence of effects while assessing the parallel trends assumption.

\subsubsection{Difference-in-differences: exploiting household panel data}
We estimate a two-way fixed effects DiD model:
\begin{align} \label{reg:dd}
O_{imt}= \alpha_s \indicator_{m \in \textit{Shaken}}+ \alpha_1 \indicator_{m \in \textit{Shaken}} \times \indicator_{t \in \textit{After}} +\mathbf{X}_{it}'\alpha_X +\indicator_t + \indicator_i + \epsilon_{imt},
\end{align}
\noindent where $O_{imt}$ denotes the outcome of interest (stated discount factor, consumption, or savings). The coefficient $\alpha_1$ captures the average treatment effect on the treated (ATT) for households exposed to shaking compared to those that experienced no ground movements. The indicator $\indicator_{m \in \textit{Shaken}}$ is one for municipalities $m$ in the shaken but not devastated treatment group; the indicator $\indicator_{t \in \textit{After}}$ is one for post-earthquake observations. We compare households in municipalities that were \textit{shaken but not devastated}, those that experienced ground motion but little or no damage, to households in areas with no recorded seismic activity. By tracking the same households before and after the earthquake, we isolate within-household changes in stated preferences attributable to visceral exposure. Year fixed effects $\indicator_t$ capture time trends common to all households, and household fixed effects $\indicator_i$ control for time-invariant unobserved heterogeneity at the household level.

A key identifying assumption for DiD is that outcomes of the treatment and control groups would have evolved in parallel in the absence of the earthquake. Parallel trends are plausible because, as we discussed in Section \ref{s:earthq}, ground motion propagated with quasi-random variation around the epicenter. This enables comparisons across households that are similar in observed characteristics but differ in seismic exposure. Table \ref{tab:stat} confirms that treated households (\textit{shaken but not devastated}) and two control groups: \textit{rest of Italy} and \textit{spatial control}, were broadly balanced on pre-earthquake observables.\footnote{While treated households are smaller and heads are more likely to have university degrees, these reflect slowly-evolving regional characteristics, for instance, the concentration of high-skilled employment in the broader commuting zone around Rome. Moreover, these differences are small in magnitude and unlikely to drive differences in post-treatment outcomes. Appendix Table \ref{tab:stat40k} shows that the observable characteristics among households living in municipalities within 40 km of the epicenter were, on average, also similar to the \textit{shaken but not devastated} group (column 1), but they have significantly lower pre-quake discount factors. This underscores a key limitation of post-event-only designs. Without pre-treatment measurements, underlying spatial variation in preferences may conflate treatment effects.} We later confirm pre-earthquake parallel trends with an event study design.

\begin{table}[h!]
 \caption{Pre-Earthquake Descriptive Statistics} \label{tab:stat}
 \footnotesize
\begin{tabular}{llccccc}
\toprule
                                                                  &  & Shaken & \textit{p-value}  & Spatial Control & \textit{p-value} & Rest of Italy   \\
                                                                                   &  & (1)    & (2)               & (3)           & (4)              & (5)            \\
\hline
\textbf{Head characteristics}     &  &        &                   &               &                  &                \\
Stated discount factor (0-100)                                    &  & 91.80  & 0.502           &     92.21                   & 0.797          & 91.94         \\

Male                                                                   &  & 0.65   &  0.023              &     0.71                       & 0.012        & 0.71          \\

University degree (dummy)                                  &  & 0.18   &   \textless{}0.001                      &   0.07           & \textless{}0.001       & 0.09          \\

Employed (dummy)                                           &  & 0.65   & 0.007           &  0.58                  & 0.031          & 0.61           \\
                                                           &  &        &                   &               &                  &                \\
\textbf{Household characteristics} &  &        &                   &               &                  &                \\
Annual household income                     &  & 27,853 &     0.099                      &   26,359                              & 0.175            &  26,762        \\

Annual savings to income ratio (0-1)       &  & 0.21   &    0.996                           &   0.21                   & 0.381      & 0.23        \\

Wealth                                       &  & 256,746 &   0.013           &    212,235                                         & 0.019      & 221,066           \\

Family size \textgreater 2 (dummy)        &  & 0.37  &    \textless{}0.001          &     0.52               &  \textless{}0.001          & 0.49           \\

Have children (dummy)                          &  & 0.20   &      0.349                                     &  0.22         & 0.091           & 0.23          \\
                                                           &  &        & \textit{}         &               & \textit{}        &                \\
 
 No. of households                                       &  & 511    & -                     & 1,874       & -                &   18,248  \\  
\bottomrule            
\end{tabular}

 \vspace{1mm}
  \hspace{1mm}
  \begin{minipage}{1.02\linewidth}

\footnotesize 
\textit{Notes}: Pre-earthquake means for households in shaken municipalities excluding L'Aquila (col. 1), spatial control ($<$125 km) (col. 3), and the rest of Italy (col. 5). P-values test equality of means with the shaken group. Stated discount factor: 2004 and 2008 SHIW waves (2008 restricted to pre-earthquake interviews). All other variables: 2002–2008 waves. Monetary values in 2010 euros. Survey weights applied.
\end{minipage}

\end{table}

After natural disasters, the most adversely affected households are often relocating, and attrition is a common concern. Our treatment, however, is exposure to seismic motion rather than to physical destruction, so relocation incentives are substantially weaker.%
 We provide formal evidence against differential attrition of treated and control in Online Appendix Tables~\ref{tab:balance_attrit} and~\ref{tab:attrit_eq}.\footnote{Using interviewer contact forms collected by Bank of Italy, we construct household-level 
attrition indicators distinguishing refusal, displacement, and death, and exclude households rotated out by survey design. The balance table shows that 
stayers and genuine attritors are broadly similar on pre-earthquake characteristics, and stated patience is virtually identical 
across the two groups (91.7 vs 91.0, p=0.206). The regression evidence confirms that earthquake exposure predicts no category of attrition except refusal, marginally, in the full sample. A placebo test using pre-earthquake attrition in the 2006-2008 window yields similarly insignificant results.}

\subsubsection{Event study: dynamics and duration of preference shocks}
To test the timing and persistence of effects, and to validate the parallel trends assumption, we estimate an event study specification:
\begin{equation} \label{eq:es}
O_{imt} = \alpha_s \indicator_{m \in \textit{Shaken}}+\sum_{t\neq 2008}\alpha_t \indicator_{m \in \textit{Shaken}}\times \indicator_{t } + X^{'}_{imt}\alpha_X +\indicator_t + \indicator_i +\epsilon_{imt}.
\end{equation}

$\alpha_t$ traces the treatment effect in each year relative to 2008.  For stated preferences, the 2008 reference group includes only
respondents interviewed before the earthquake. While stated preferences are available in only two post-earthquake waves (2010 and 2012), the consumption and savings analysis covers all waves from 2002 to 2016. The key idea is that if preference shocks are persistent, they should manifest as lasting behavioral shifts. This also allows us to establish a longer parallel pre-trend.

 \section{Empirical Results}\label{s:result}
\subsection{Stated Time Preferences}
\subsubsection{Distributional Evidence}
\begin{figure}[h!]
\centering 
\subfigure[Shaken Treatment]{\label{fig:a}
\includegraphics[width=68mm]{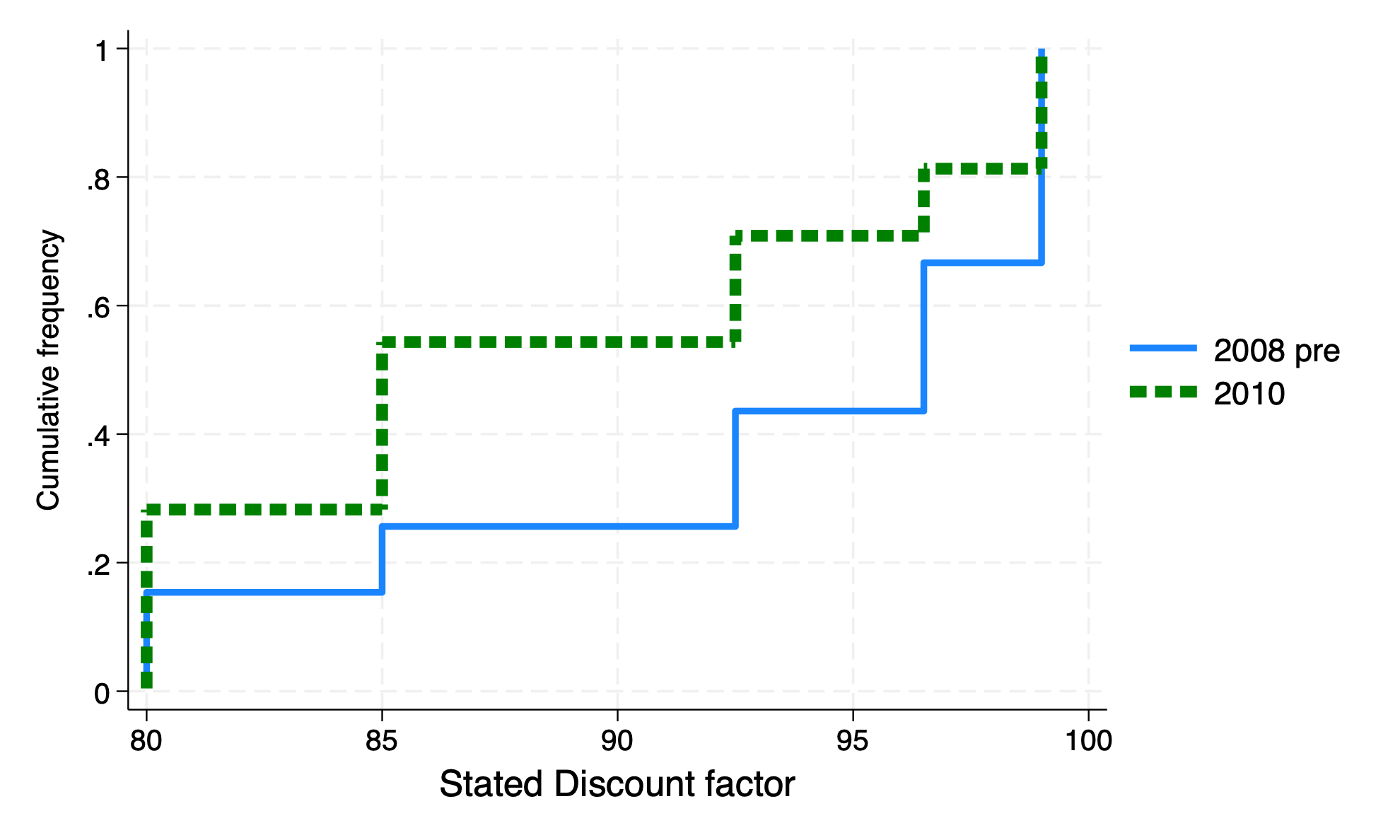}}
\subfigure[Control]{\label{fig:b}
\includegraphics[width=68mm]{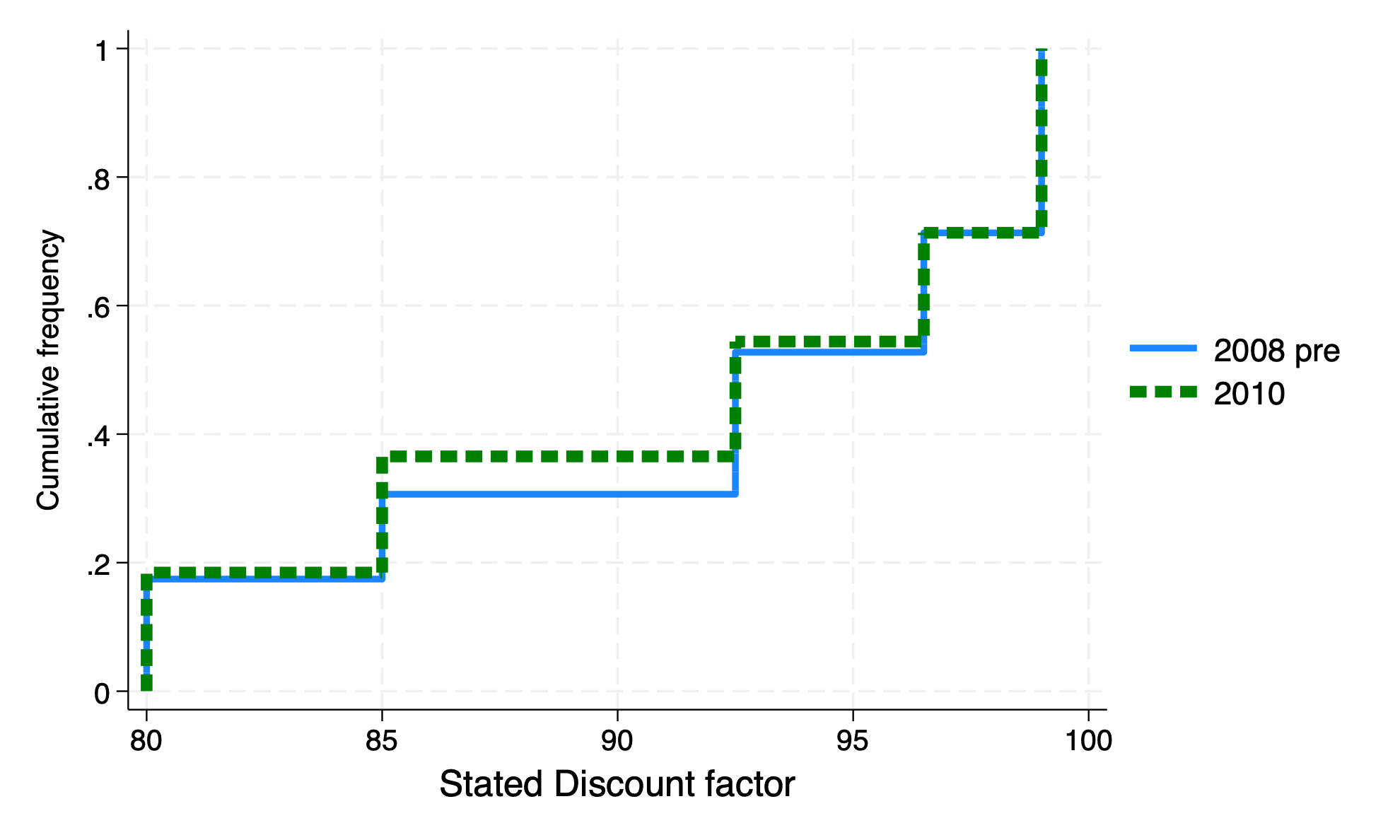}}
\caption{Stated Discount Factor Before and After the Earthquake. \textit{Notes}: The figure plots the cumulative distribution of stated discount factor for individuals living in shaken and control municipalities.}\label{fig:df_cdf}      
\end{figure}

Comparing cumulative distribution functions of stated time preferences for the treated and control groups right before and right after the earthquake (Figure~\ref{fig:df_cdf}), we can see that the entire distribution shifted leftward. This indicates that the decline in patience was not driven by a few respondents but was pervasive across the sample.\footnote{Prior to the earthquake, the distributions of stated time preferences were statistically indistinguishable across treatment and control groups ($\chi^2(4) = 1.10$, $p = 0.89$). Following the earthquake, the distributions are significantly different ($\chi^2(4) = 28.67$, $p < 0.001$).}
\subsubsection{Difference-in-Differences Estimates}
We report the DiD estimates in Table~\ref{tab:dd_tp}.\footnote{Because treatment varies across a small number of municipalities, we also compute randomization-inference p-values, permuting treatment across municipalities \citep{conley2011inference,mackinnon2020randomization}. For stated patience, $p = 0.010$; the consumption response is less precisely estimated under this procedure ($p = 0.41$ against the spatial control group).} Panel A uses the national unshaken sample as the control group, while panel B uses the spatial control group. 
The results show a decline of approximately 5 percentage points in the stated discount factor for the treated group, equivalent to a shift from the median to approximately the 25th percentile of the national distribution.\footnote{The DiD estimates exceed the OLS estimates reported in Table~\ref{tab:nakamura}, indicating that the conventional design based solely on post-disaster elicitation of stated preferences understates the treatment effect, as unobserved heterogeneity attenuates the estimated response.}

The near-identical coefficient with and without income (column 1 and 2) indicates that income is not absorbing a meaningful part of the treatment response. The treatment effect remains stable after adding demographic controls (column 3), confirming that shifts in household composition do not drive the result. Columns 4--6 then sequentially add variables capturing changing financial conditions: liquid assets, wealth, and a proxy for being credit constrained.\footnote{A household is coded as credit-constrained if it was refused credit, granted only part of the amount requested, or discouraged from applying in anticipation of rejection.} The stability of the coefficient across these columns addresses the concern that the preference shift is an artifact of tightened financial conditions.

Column 7 additionally controls for HTM status, using an indicator of hand-to-mouth status: liquid wealth falls short of one week's income (\citet{kaplan2014wealthy}) , or financial wealth no greater than half of annual earnings (\citet{Zeldes} and \citet{Hitoshi21}). The results are unchanged, indicating that becoming HTM is not driving the stated patience results.

Taken together, the progression across columns demonstrates that the decline in stated patience is not explained by pre-existing group differences, demographic change, income shocks, wealth effects, credit frictions, or liquidity constraints, but instead reflects a direct behavioral response to earthquake exposure. 

\subsubsection{Spatial Control}\label{subsec:results-spatial}
A remaining concern is that even undamaged households in our treatment group can be impacted indirectly by the earthquake through work, family, or general equilibrium effects. For example, disrupted labor markets,  social obligations, or price shocks might impact the estimated treatment effect. To address this, we restrict comparison to a spatial control group: households in unshaken municipalities within 125 kilometers of the epicenter, the maximum range of recorded ground 
motion. As formalized in Section~\ref{s:framew}, spatial control has to satisfy three conditions. Treated and spatial control households should be balanced on pre-earthquake observables, which we verify in Table~\ref{tab:stat} (condition \textit{i.}). Treated and spatial control households have parallel disposable-income trajectories before and after the earthquake (Appendix Figure~\ref{fig:event125_y}; condition \textit{ii.}). Income in the region fell, and Abruzzo's regional GDP contracted by 6\% in 2009. However, disposable income in the treated and spatial control groups changed in a parallel fashion.  Finally, prices may increase near the damaged area; therefore, treated and control households should be at similar distances from the damaged area on average, which we verify in Appendix Table~\ref{tab:dist_balance}. We do further robustness tests on prices, and distance to the epicenter in the consumption section \ref{sec:cons}.%

Estimated treatment effects reported in Panel B are stable across specifications and close to those obtained with the national control group, indicating that regional spillovers common to treated and spatial-control households do not drive the decline in stated patience.

\begin{table}
\caption{Difference-in-Differences Estimates: Stated Time Preference}
\label{tab:dd_tp}
\scriptsize
\setlength{\tabcolsep}{4pt}
\resizebox{\textwidth}{!}{%
\begin{tabular}{l*{7}{c}}
\toprule
                    
                    &\multicolumn{1}{c}{(1)}&\multicolumn{1}{c}{(2)}&\multicolumn{1}{c}{(3)}&\multicolumn{1}{c}{(4)}&\multicolumn{1}{c}{(5)}&\multicolumn{1}{c}{(6)}&\multicolumn{1}{c}{(7)}\\
\hline
\textit{Panel A: Rest of Italy}
&&&&&&&\\
Shaken x After      &      -5.468&      -5.413&      -5.536&      -5.536&      -5.535&      -5.531&      -5.540\\
                    &     (0.595)         &     (0.596)         &     (0.604)         &     (0.605)         &     (0.604)         &     (0.605)         &     (0.607)         \\
&&&&&&&\\
Mean of dep. var.   &        91.9         &        91.9         &        92.1         &        92.1         &        92.1         &        92.1         &        92.1         \\
Observations        &       14,413         &       14,413         &       13,688         &       13,688         &       13,688         &       13,688         &       13,688         \\
R-squared           &       0.520         &       0.520         &       0.522         &       0.522         &       0.522         &       0.522         &       0.522         \\
&&&&&&&\\
 
\textit{Panel B: Spatial Control}
&&&&&&&\\
Shaken x After      &      -6.047&      -6.061&      -6.403&      -6.121&      -6.411&      -6.384&      -6.410\\
                    &     (0.854)         &     (0.848)         &     (0.892)         &     (0.927)         &     (0.891)         &     (0.889)         &     (0.884)         \\
&&&&&&&\\
Mean of dep. var.   &        91.9         &        91.9         &        91.9         &        91.9         &        91.9         &        91.9         &        91.9         \\
Observations        &        1,517         &        1,517         &        1,464         &        1,464         &        1,464         &        1,464         &        1,464         \\
R-squared           &       0.556         &       0.556         &       0.566         &       0.569         &       0.566         &       0.569         &       0.567         \\
&&&&&&&\\
Income            & no  & yes & yes & yes & yes & yes & yes \\
Controls          & no  & no  & yes & yes & yes & yes & yes \\
Liquid asset      & no  & no  & no  & yes & no  & no  & no  \\
Wealth            & no  & no  & no  & no  & yes & yes & yes \\
Credit constraint & no  & no  & no  & no  & no  & yes & no  \\
HTM status        & no  & no  & no  & no  & no  & no  & yes \\
\hline
\end{tabular}
}
\legend{Dependent variable: stated discount factor (0--100). All columns report difference-in-differences estimates and include household and year fixed effects. Controls include the household head's age and education, family size, and the presence of children. Data are from the SHIW waves of 2004, 2008, 2010, and 2012. Standard errors, clustered at the initial municipality level, are reported in parentheses.}
\end{table}

\subsection{Effects on Consumption}\label{sec:cons}
The second condition for identifying an impatience shock is that consumption should increase. We test this with non-durable consumption as our primary outcome, since durable purchases are lumpy and infrequent. We restrict the sample to non-HTM households,\footnote{HTM households constitute approximately 28 percent of the sample.}  who have the means to increase consumption (see discussion in  Section \ref{s:framew}), and apply standard life-cycle age restrictions 25-70.

Nondurable consumption rises by about €2,100 per year (€175 per month), approximately 11 percent of the pre-treatment mean (Table~\ref{tab:dd_cons}).\footnote{€175 per month, at 2010 prices, roughly corresponds to the cost of one restaurant dinner for two per week.} 
The estimate is larger when conditioning on income (column 2), because income has declined in the treatment group, which counteracts the consumption increase. Controlling for household covariates leaves the result unchanged, showing that the consumption increase is not due to household compositional changes. The consumption increase also holds when comparing to households with similar liquid assets (Column 4), wealth (column 5), credit constraint (column 6), or HTM status (column 7).%
 This rules out the possibility that the consumption increase reflects changing financial conditions rather than preference shifts. Panel B confirms similar results against the spatial control group. Because spatial control households experienced an income decline similar
to that of treated households, this comparison does not confound the
consumption response with the treated group's income loss, and the
estimates are somewhat larger. It also nets out indirect and general
equilibrium effects common to the region.

\begin{table}
\small
\caption{Short-Run Effects on Annual Non-Durable Consumption: Non-Hand-to-Mouth Households}
\label{tab:dd_cons}
\begin{tabular}{l*{7}{c}}
\toprule
                    &\multicolumn{1}{c}{(1)}&\multicolumn{1}{c}{(2)}&\multicolumn{1}{c}{(3)}&\multicolumn{1}{c}{(4)}&\multicolumn{1}{c}{(5)}&\multicolumn{1}{c}{(6)}&\multicolumn{1}{c}{(7)}\\
\midrule
\textit{Panel A: Rest of Italy} &&&&&&& \\
Shaken $\times$ After & 1412.9 & 2108.2 & 2121.8 & 2139.9 & 2128.8 & 2107.3 & 2102.1 \\
                      & (356.1) & (405.9) & (406.4) & (399.9) & (407.0) & (402.2) & (397.8) \\
&&&&&&& \\
Mean of dep. var.     & 19,149 & 19,149 & 19,165 & 19,165 & 19,165 & 19,165 & 19,165 \\
Observations          & 11,590 & 11,590 & 11,537 & 11,537 & 11,537 & 11,537 & 11,537 \\
R-squared             & 0.694  & 0.728  & 0.730  & 0.730  & 0.730  & 0.730  & 0.730  \\
&&&&&&& \\
\textit{Panel B: Spatial Control} &&&&&&& \\
Shaken $\times$ After & 1694.8 & 2302.0 & 2361.2 & 1920.6 & 2635.2 & 2643.6 & 2639.4 \\
                      & (914.7) & (993.8) & (1142.8) & (1031.4) & (992.4) & (991.5) & (979.6) \\
&&&&&&& \\
Mean of dep. var.     & 19,658 & 19,658 & 19,667 & 19,667 & 19,667 & 19,667 & 19,667 \\
Observations          & 1,277  & 1,277  & 1,271  & 1,271  & 1,271  & 1,271  & 1,271  \\
R-squared             & 0.668  & 0.712  & 0.714  & 0.717  & 0.717  & 0.717  & 0.717  \\
&&&&&&& \\
Income            & no  & yes & yes & yes & yes & yes & yes \\
Controls          & no  & no  & yes & yes & yes & yes & yes \\
Liquid asset      & no  & no  & no  & yes & no  & no  & no  \\
Wealth            & no  & no  & no  & no  & yes & yes & yes \\
Credit constraint & no  & no  & no  & no  & no  & yes & yes \\
HTM status        & no  & no  & no  & no  & no  & no  & yes \\
\bottomrule
\end{tabular}
\legend{Dependent variable: annual non-durable consumption (2010 euros), excluding imputed rent. All columns report difference-in-differences estimates (equation~\ref{reg:dd}) and include household and year fixed effects. Controls comprise the household head's age and education, family size, and the presence of children. Data are from the SHIW waves of 2002 to 2012. The sample consists of household heads aged 25 to 90 at the time of the earthquake and aged 25 to 70 at the time of the survey, classified as non-hand-to-mouth in all pre-earthquake waves. Column 7 additionally controls for contemporaneous hand-to-mouth status. Standard errors, clustered at the initial-municipality level, are reported in parentheses.}
\end{table}

To benchmark the consumption change, we estimate consumption decline from an unemployment shock. Using pre-earthquake SHIW data, we estimate an MPC of 21 percent by instrumenting income with job loss shocks, following \citet{patterson2023matching} and \citet{hanksson} (see Appendix~\ref{appendix:empshock} for details). In the first year of unemployment, the average household loses approximately €9{,}500 annually (€792 per month),  implying a consumption decline of roughly €166 per month (Appendix Figure~\ref{fig:event_unemp}). The earthquake-induced increase of €175 per month is of comparable order, but in the opposite direction, suggesting that the consumption change is economically substantial.

We use the richness of our data to examine indirect earthquake effects that could drive the consumption increase in the Online Appendix. First, consumption estimates could have been affected by increases in rental prices near L'Aquila.  However, consumption results are robust, whether or not rental prices or imputed rent are included in the consumption measure; see Figure \ref{fig:event_consrent}. 
While individual consumption categories are too sparse to analyze formally, the raw data show no increase in extraordinary maintenance or household size (excluding newborns), so the consumption increase does not reflect
renovations or sheltering relatives (Figure \ref{fig:alt_stories}). 
We also perform an indirect test to examine the role of closeness to the damaged areas, and exclude households that are  within 15~km of damaged areas; the consumption increase is not driven by these households that live very close to the damage, see Figure~\ref{fig:event_cons_robust_distance}. 

Stated patience declines while nondurable consumption rises, satisfying
both conditions of our framework (Section~\ref{s:framew}). Together, these
results are consistent with an impatience time preference shock.\footnote{Although SHIW lacks detailed consumption categories, we can examine durable consumption, which also increased (Table~\ref{tab:dd_c_rob}).}  Next we examine whether these effects persisted over time.

\subsection{Dynamics}\label{subsec:event}
We use an event study design to assess how rapidly the time preference shock emerged and whether it persisted long enough to generate persistent reallocations in savings. 

\subsubsection{Immediate and Persistent Changes in Stated Preferences}
\begin{figure}[h]
    \centering
    \includegraphics[width=0.8\linewidth]{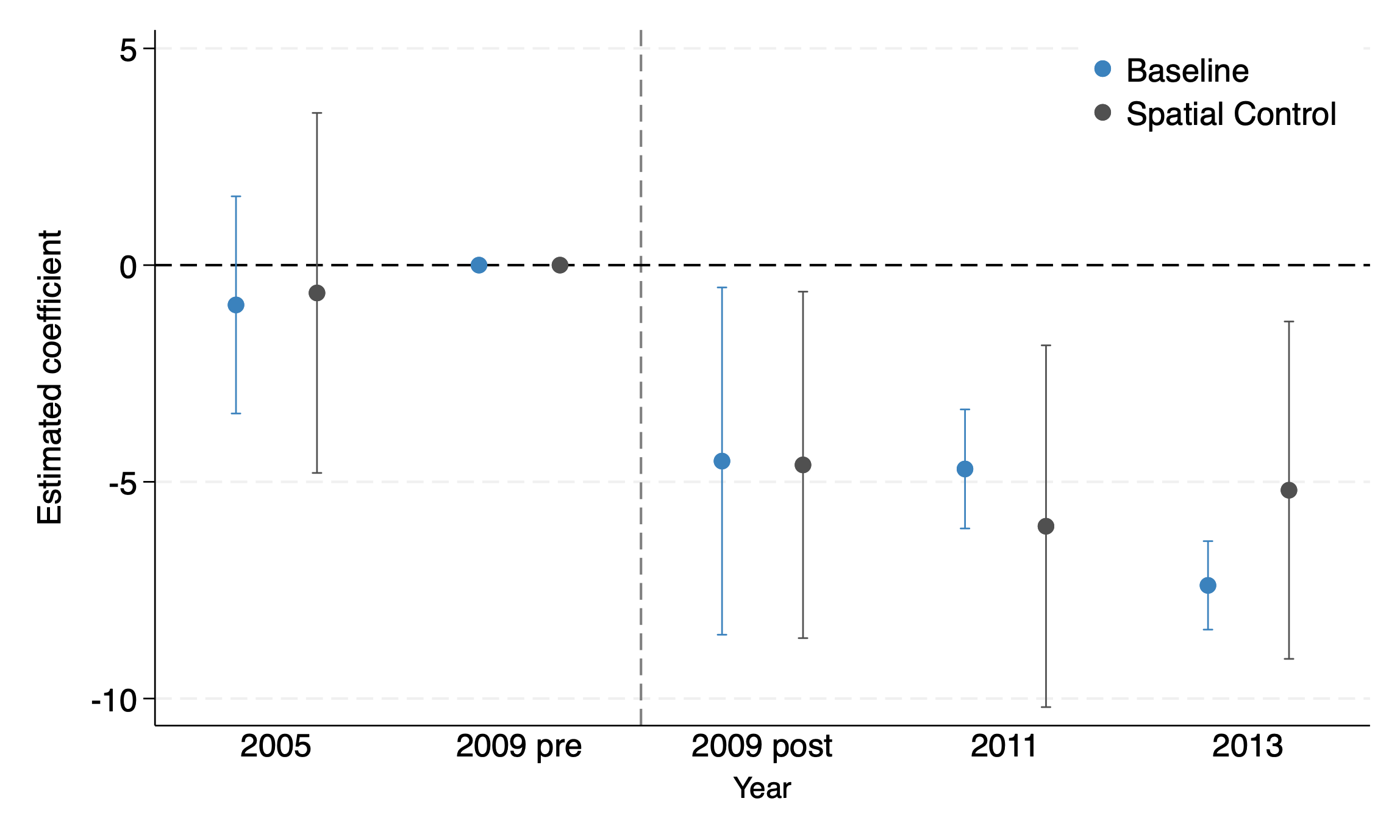}
    \caption{Effects on Stated Discount Factor - Event Study. \label{fig:event_tp}} \legend{Event study estimates of equation \ref{eq:es} (TWFE) on  stated discount factors. Reference period:  2008 wave, interviews before 6 April 2009. Baseline include all analysis-sample without controlling for risk aversion. Blue: control group from unshaken households in the rest of Italy. Grey: spatial control.}
    
\end{figure}

As before, we first present results on the whole sample, without life-cycle restrictions. Figure \ref{fig:event_tp} shows that stated patience fell sharply and immediately: respondents interviewed within weeks of the earthquake in 2009 already report significantly lower stated discount factors
relative to respondents interviewed in 2009 prior to the earthquake. Stated discount factors in the treatment group remain low even in the 2011 and 2013 interviews.  
The same pattern holds when we compare treatment to the spatial control group, confirming that regional economic spillovers do not drive the results.
SHIW elicited time preferences again in 2020, and in Appendix Figure~\ref{fig:event_df2020}  we show that those who were still surveyed from the treated group on average still had low stated patience. This suggests that the time preference shock could be very persistent, but the sample size gets too small to make a definite conclusion.\footnote{The stated discount factor question was dropped from the SHIW after 2012 and reintroduced only in the 2020 wave. We extend the analysis to this wave despite one limitation: the SHIW's rotating panel design, compounded by attrition, leaves only 17 treated households observed in 2020. With this caveat, Appendix Figure~\ref{fig:event_df2020} shows that both the event-study estimate and a cross-sectional specification with region fixed effects yield negative point estimates of similar magnitude to the 2012 coefficient, providing suggestive evidence that the decline in stated patience persisted more than a decade after the earthquake.}  The lack of pretrends visible in Figure~\ref{fig:event_tp} is consistent with parallel trends.

\subsubsection{Persistent Effects on Consumption and Savings}
If the time preference decline is persistent, it should generate a persistent reallocation in consumption and savings. We extend our sample and follow households seven years post-earthquake, four years beyond the last stated-preference observation, to examine this.
 \begin{figure}[h!]
 \centering
\includegraphics[width=0.85\linewidth]{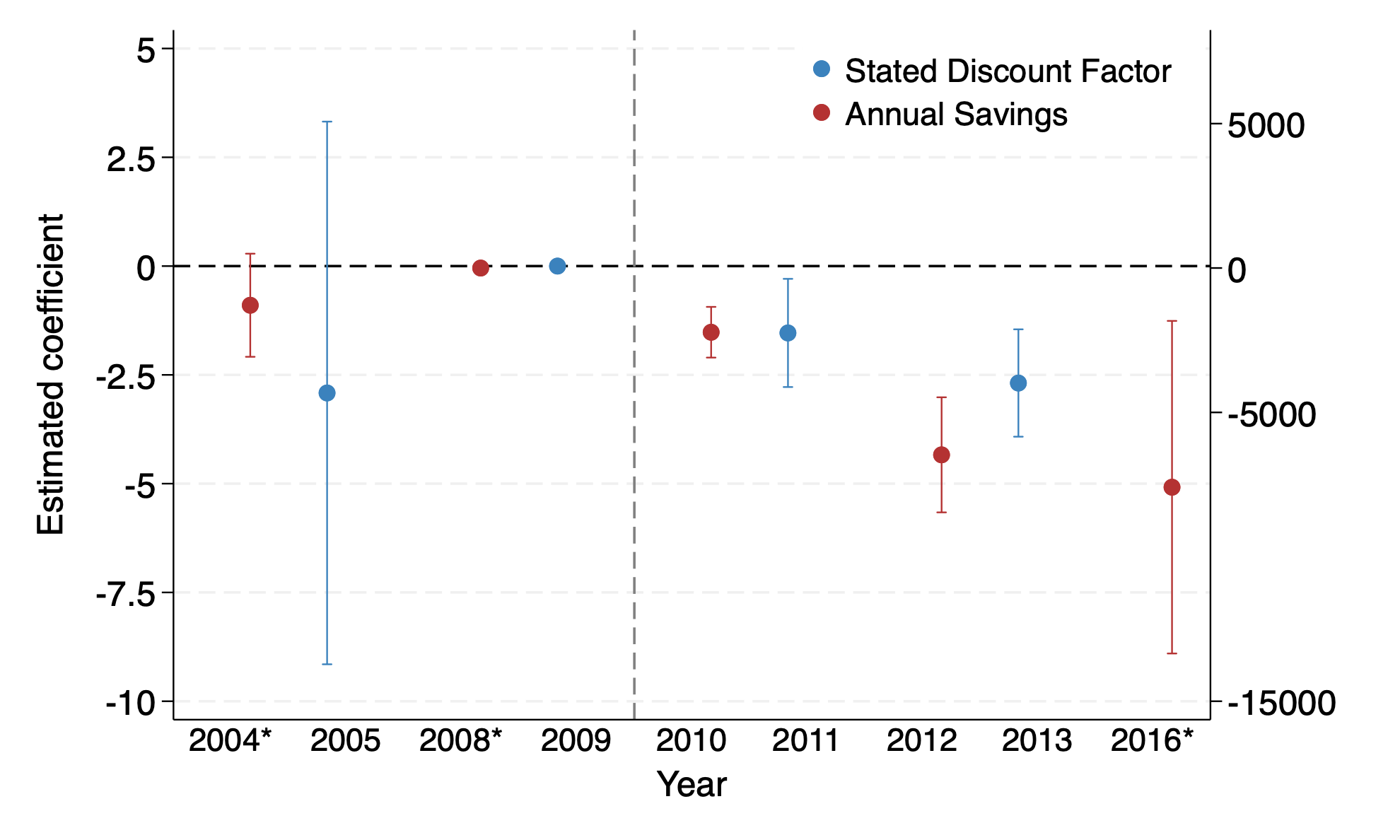}
 \caption{Long Term Effects on Savings and Stated Patience (Unconstrained Households) - Event Study. \label{fig:event-s}}
\legend{Event study estimates of earthquake exposure on stated discount factor and annual savings. Sample: non-HTM in all pre-earthquake periods, heads aged 25–70 at the time of the survey. Due to limited sample sizes, 2002–2004, 2006–2008, and 2014–2016 are pooled for savings (red); robustness without pooling shown in Figure \ref{fig:event_poolorisav}. For consumption results see Figure~\ref{fig:event_cons}. Left axis: stated discount factor coefficients; right axis: savings coefficient(2010 euros). $\sym{*}$ indicates pooled years.}  
\end{figure} 

Among working-age, non-HTM households prior to the earthquake, savings fell persistently following the earthquake, as households kept saving less from their disposable income compared to similar households in the control group (Figure~\ref{fig:event-s}).\footnote{Confidence intervals widen in later years as panel attrition reduces sample size. To improve precision at the panel's tails, we pool adjacent survey waves into two-wave bins (2002–2004, 2006–2008 and 2014–2016), keeping 2010 and 2012 as separate periods.} 
Non-durable consumption out of disposable income increases at the same time, before the difference becomes statistically insignificant, as households are not able to sustain an increased consumption conditional on income over seven years. (Appendix Figure~\ref{fig:event_cons}).\footnote{Durable consumption also rises in the short run but attenuates by 2016; see Appendix Figure~\ref{fig:event_cons}.} In Figure~\ref{fig:event-s} we also reconfirm that stated patience also declined in this subsample. 

Taken together, the persistent reduction in savings among households with the financial means to maintain pre-shock plans is consistent with a lasting shock to time preferences.

\subsubsection{Financing the Consumption Increase}
What are the long-run consequences on household financial resources? We examine this by following \citet{georgarakos2024} and \citet{coibionetal2024} and estimate how changes in long-run financial assets depends on stated patience changes:

\begin{align} \label{reg:ecb}
F_{imT}=  \gamma_t + \gamma_m + \alpha_1 \Delta \tilde{\beta}^{Post}_{it} +  \alpha_2 \tilde{\beta}^{Pre}_{it} +\mathbf{X}_{it}'\alpha_X  + \epsilon_{imt},
\end{align}
where $F_{imT}$ denotes the household's rank in the distribution of liquid assets, broader financial holdings, or bank liabilities in the last waves in our sample, five to seven years after the earthquake (we average 2014 and 2016).\footnote{$\Delta \tilde{\beta}^{Post}_{it}$ is computed from the same households used to estimate the DiD. Although the stated discount factor is a noisy measure of the true underlying discount factor, to the extent that measurement error is classical, the coefficients are attenuated toward zero, and the estimates in Table~\ref{tab:portfo} are therefore conservative.}
    
    Appendix Table~\ref{tab:portfo} shows a clear pattern of financing increased consumption through dissaving and increased borrowing. As an example, prior to the earthquake, treated households ranked at approximately the $57$th percentile of the national liquid wealth distribution. By 2014-2016  they moved down by $2.6$ percentile points, to approximately the 54th percentile.\footnote{For this back-of-the-envelope calculation we scale the coefficient in column~(1) of Table~\ref{tab:portfo} by the DiD estimate of the earthquake's effect on patience (approximately $-5$ discount factor units, Table~\ref{tab:dd_tp}).} The change in stated time preferences mediates the decline in liquid assets: those with a bigger decline in stated patience decreased their liquid assets more.
    Households used not only their savings, but also borrowed more to finance the consumption increase. Those with a larger post-earthquake decline in stated patience carried higher bank liabilities by 2014-2016.
    Controlling for pre-earthquake stated patience levels leaves the estimates unchanged, confirming that earthquake-induced shifts in preferences, rather than stable individual heterogeneity, drive these changes in financial choices. 

Together, the evidence on persistent stated impatience, persistent savings decline, and deteriorating balance sheets suggest that the earthquake generated a lasting shift in latent time preferences with material consequences for household financial buffers.  
\section{Mechanisms}\label{s:mechanism}

How can a traumatic earthquake experience induce such persistent changes in time preferences? We consider various explanations. Though some are difficult to separate from genuine time preference change, since they also operate through the Euler equation, the richness of our data allows us to bound their contribution.

\subsection{Curvature of the utility function}
A large empirical literature demonstrates that risk preferences can change after natural disasters and other traumatic events (\citet{Hanaoka_et_al_Japanriskpref}; \citet{cheong2022natural}; \citet{quiroga2024revealing}).  Because elicited discount rates are overstated when utility curvature is ignored \citep{AndersenHarrisonLauRutstrom_ECTA08_risk_time}, an increase in risk aversion would lower measured patience even if time
preferences were unchanged.

Appendix Figure~\ref{fig:event_rp} demonstrates, using a recurring investment risk-preference question (see Appendix~\ref{appx:data}), that risk aversion did increase in our treated group, but returned to pre-earthquake values after two waves.  However, increased risk aversion is difficult to square with the time pattern of savings we observe. Higher risk aversion should raise precautionary savings, while the pattern in the data is the reverse. The saving decline persisted for seven years, well after risk
aversion had returned to pre-earthquake levels.

We also estimate how time preferences would have changed if we correct for changing risk preferences, following \citet{Hitoshi21}. For this correction, we need an incentive-compatible lottery measure of the Arrow–Pratt risk aversion, because the investment risk-preference question may partly reflect beliefs about returns rather than pure risk aversion. We have such a risk-preference measure from 2010. We use this measure combined with the investment risk-preference measure to impute Arrow-Pratt risk preference measures for all waves (see details in Appendix~\ref{subsec:app_timerisk}). We find that even with the new, risk-aversion-corrected measure, stated patience declines, though the magnitude is smaller  (Table~\ref{tab:patience_corr}). 
As a final test, we also control for risk aversion directly in the DiD estimation, which gives the same result: stated patience declines (Table~\ref{tab:patience_corr}, column 4). None of these support the mechanism that changing risk preferences drive our results. 

\subsection{Uncertainty, beliefs}
Disasters raise both objective and perceived uncertainty \citep{guiso2008trusting-stockmkt, baker2024usingsisasters, Gallagher_AEJApp14}.  Income and employment uncertainty likely rose throughout the region. Provided it rose similarly in treated and spatial control groups, the spatial control estimates net it out. Subjective uncertainty may nonetheless have risen more in the treated group, for whom the earthquake experience was more salient. However, under standard preferences, greater uncertainty should raise precautionary saving and dampen current consumption \citep{Carroll1997} - the opposite of what we observe. 

Changing beliefs about future income or returns could also produce a decline in savings, but the survey data suggest otherwise. Treated and control households hold similar income expectations after the earthquake.\footnote{SHIW asks whether respondents expect household income to increase more than prices from 2010 onwards. The average answer is not statistically different $p=0.69$ in treated (1.59) and control (1.60). Because the question is not asked before the earthquake, we cannot rule out pre-existing level differences.} Treated households revised their interest rate and stock market return expectations upward (see Table~\ref{tab:dd_exp}), but this should not impact their intertemporal choices. Their elasticity of intertemporal substitution is about 1, so the income and substitution effects of higher expected returns cancel out.\footnote{The treated group's imputed Arrow-Pratt risk aversion after the earthquake is 1.02.} 

Treated households may also have lowered their subjective survival probability. In a mortality-adjusted Euler equation, a decline in survival probability $\delta_{i,t}$ is observationally equivalent to a decline in $\beta_{i,t}$:
\begin{align}\label{eq:euler-mortality}
u'(C_{i,t})=\delta_{i,t}\beta_{i,t}E_t\left[(1+r)\,u'(C_{i,t+1})\right].
\end{align}
To bound this channel, we ask how much life expectancy would have to fall if our estimated five percentage point decline in stated time preferences were entirely driven by lower survival probability. A 40-year-old in 2010 had a remaining life expectancy of 42 years; a five percentage point decline in annual survival probabilities would cut this to 21 years.\footnote{Own calculations based on data from the Italian National Institute of Statistics (ISTAT).} This magnitude seems unrealistic. It is also a high reduction compared to  \citet{FalkHermleSunde19longevity}, who find that a five percentage point lower stated discount factor corresponds to a 10-year shorter subjective life expectancy. Longevity beliefs therefore cannot account for the entire reduction in time preferences.

\citet{Gallagher_AEJApp14} highlights that increased disaster risk beliefs could weaken incentives to save, and  explain increased flood-insurance take-up after floods. We find no such response. Home insurance coverage among treated households remains low, at about 12 percent, which suggests that perceived disaster risk did not rise enough to change insurance behavior.\footnote{Low coverage may partly reflect moral hazard, as the state often rebuilds private homes. Private insurance nonetheless has value. During our 2024 visit to L'Aquila, we observed that reconstruction was substantially faster for insured households.}

\subsection{Psychological stress}
Our findings corroborate recent psychological work arguing that stress can make near-term outcomes more salient and shorten the effective planning horizon \citep[see review in ][]{Haushofer2014PsychologyPoverty}.
Experiments show that affective priming and pharmacological stress interventions reduce patience, particularly when the stressor is salient. These effects tend to fade unless the stress is chronically reinforced \citep{psych_lerner2013financial, psych_ifcher2011happinessAER, Cornelisse2013HydrocortisoneIntertemporalChoice}.\footnote{Salience refers to the disproportionate allocation of attention to certain aspects of the environment \citep{taylor1982stalking}. \citet{bordalo2012salience} develop a model of attention allocation in which shifts in salience generate systematic instability in risk preferences.}

The L'Aquila earthquake generated such salient and reinforced psychological stress. The earthquake struck in the middle of the night, waking people from their sleep, a profound form of shock that instantly activates the body's primitive fight-or-flight response. Many in L'Aquila perished, including children in a collapsed dormitory, evoking the feeling that it could have been me.\footnote{Anecdotal evidence suggests that many residents slept outside on the night of the earthquake, for example, in cars, fearful of aftershocks. The following days, some slept with unlocked doors for an easier escape. Even those living further away, in Rome, mentioned this to us in our conversations.}  
 Several reminders likely amplified the impact of initial trauma, consistent with psychological findings on chronic environmental cues \citep{lazarus1984stress}.
 The trial and conviction of public officials and geologists drew national media coverage for years \citep{cartlidge2016Science-trial}. Reconstruction was slow and marred by corruption.  Even during our 2019 visit, ten years after the earthquake, the tension was palpable. We saw residents use a scientific conference to confront local authorities over delays in rebuilding primary schools.\footnote{At the Silk Cities Conference held in L'Aquila on July 12, 2019, residents voiced frustration that, ten years after the earthquake, children were still studying in temporary buildings. Though not formally invited, they attended a keynote by the Department of Civil Protection and protested during the Q\&A.}

\begin{figure}[h!] 
\centering
\subfigure[Patience change by age group]{\label{fig:a-het}
    \includegraphics[width=65mm]{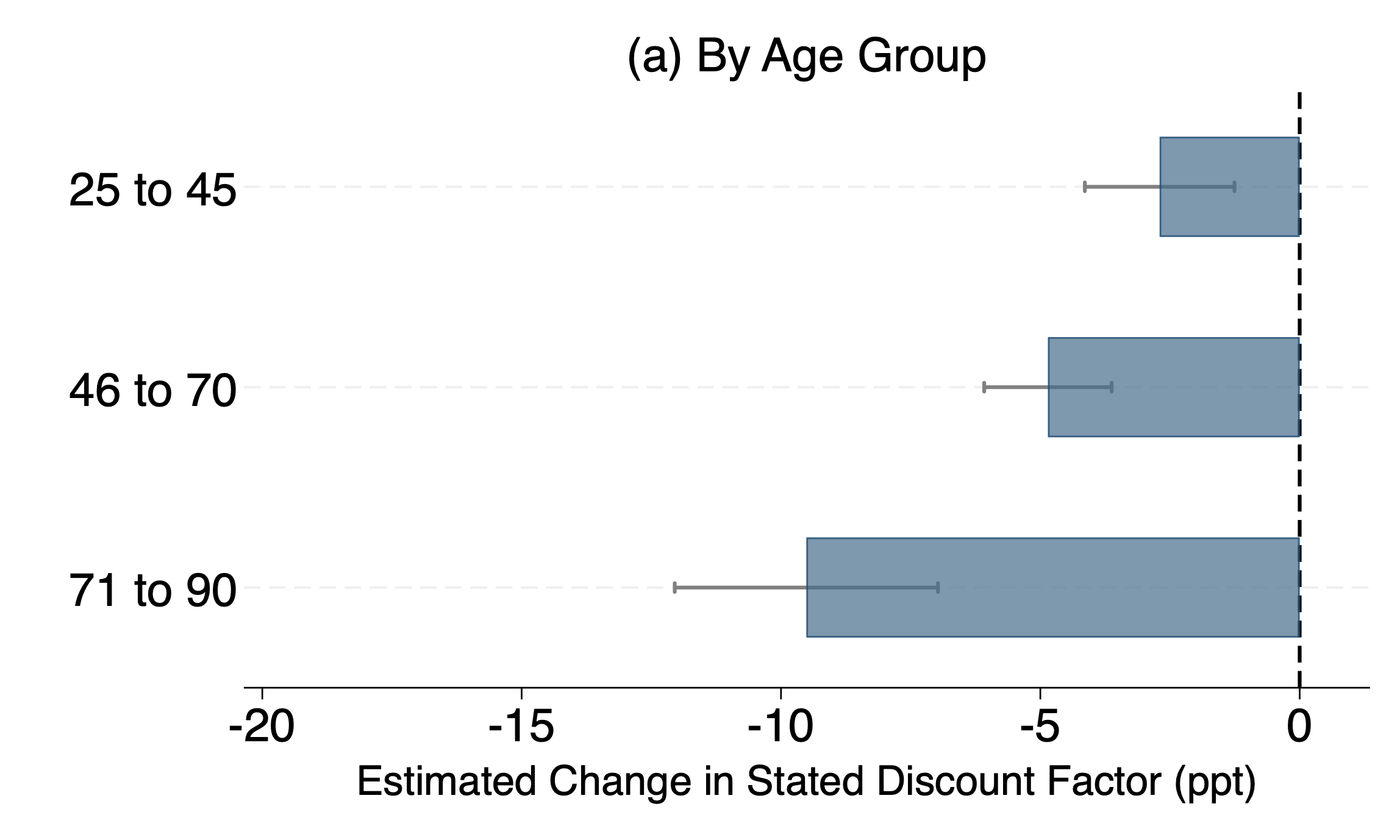}}
    \subfigure[Patience change by subjective health]{\label{fig:b-het}
    \includegraphics[width=65mm]{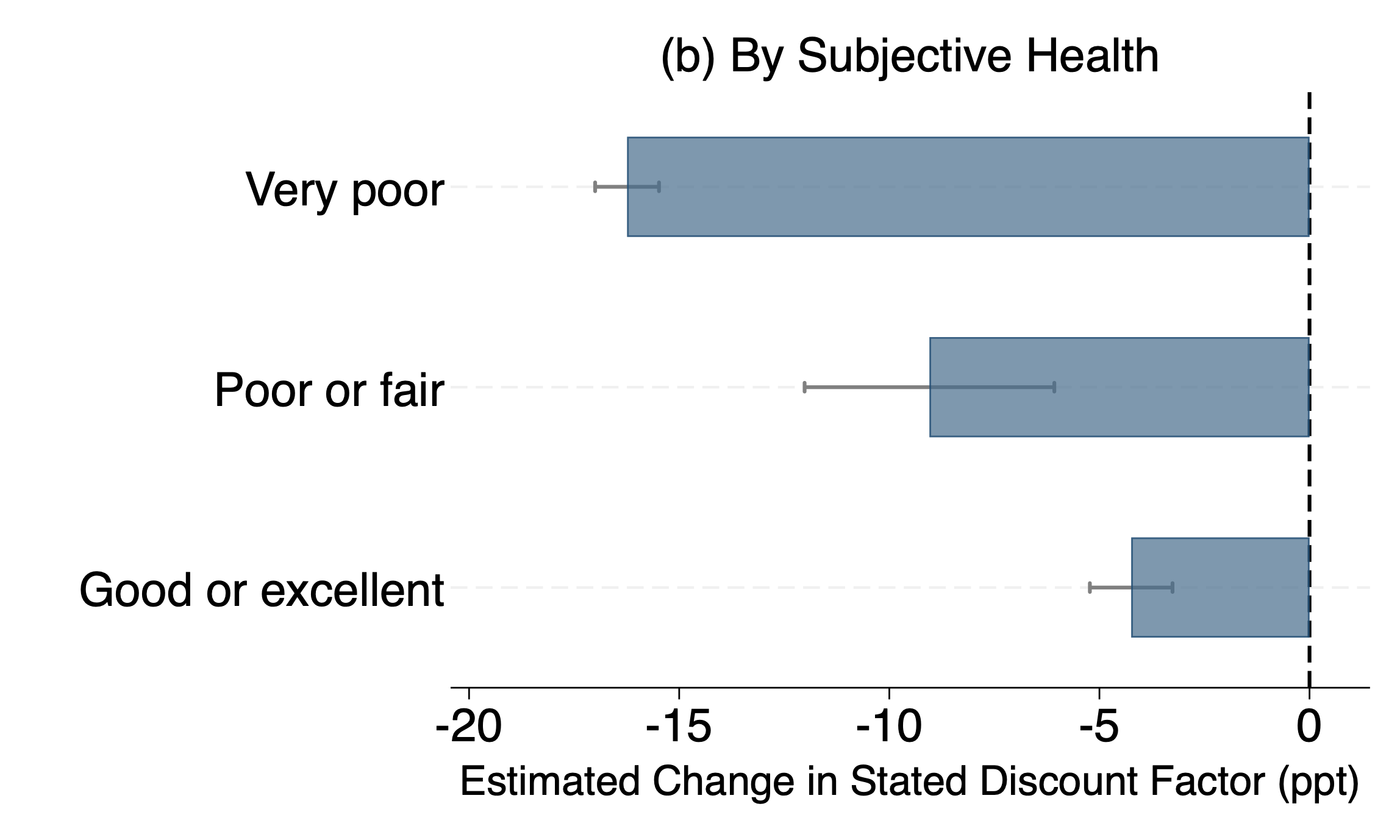}}
	\caption{Heterogeneous Preference Changes by Age and Health. \textit{Notes}: coefficients of interaction between the treatment indicator \textit{Shaken} $\times$ \textit{After} and an indicator variable for each of the three age or self-assessed health bins at the time of the L'Aquila earthquake. Capped bars: 95\% confidence intervals (see Table \ref{tab:hetero}).}
	\label{fig:patience_age_health}
\end{figure}

Although we lack direct measures of stress, we document corroborating evidence. First, the groups that changed their stated patience the most came disproportionately from groups more vulnerable to stress. Medical studies documented widespread psychological distress and stress-related sleep disorders that persisted for years both in L'Aquila and surrounding undamaged municipalities. Older individuals and those with pre-existing health problems had more sleep disturbances and higher antidepressant prescription rates (\citealt{Tempesta2013sleeping, Rossi2011Feb}).
 Stated patience declined most sharply among these exact groups: the old, aged 71 and above (Figure~\ref {fig:a-het}) and those who reported poor health before the earthquake (Figure~\ref{fig:b-het}). Second, the earthquake took a toll on health: treated households report a significant decline in self-assessed health after the earthquake (Table~\ref{tab:health}). To gauge its size, we compare it with the cross-sectional age gradient in self-assessed health, estimated in a specification with municipality rather than household fixed effects: the decline is comparable to about ten years of aging. This interpretation is consistent with experimental evidence that psychological states shift intertemporal choices when economic constraints are held fixed \citep{BartosBauerChytilovaLevely2021}. We document a similar response to a naturally occurring shock, and show that it persists for years and carries over to consumption and wealth. A psychological channel has also been documented for risk preferences:
exposure to violence raises risk aversion, more so for those with more severe trauma symptoms \citep{moya2018violence_mentalhealth} or when fear is made salient \citep{CallenAER14_afghanistan}.

\subsection{Robustness}\label{subsec:results-rob}
Online Appendix Table~\ref{tab:dd_rob3} assesses the sensitivity of the stated-patience results to sample composition, measurement, and confounding shocks. Each check is reported against both control groups: the rest of Italy (Columns 1--3) and the spatial control (Columns 4--5). 

We first address concerns about the stated time-preference measure itself. Our baseline treats the respondent's stated time preference as representative of the household, but the respondent can change across waves and different respondents may report different preferences. Column 1 (whole Italy) and Column 4 (spatial control) restrict the sample to households with a stable respondent across waves; the estimated decline in stated patience remains negative and significant, indicating that the effect is not driven by within-household respondent changes.

We probe two potential confounds specific to our setting. Rome, which is atypical in both size and administrative status, is included in the treatment group; Columns 2 and 5 show that excluding Rome leaves the point estimates virtually unchanged, although less precisely estimated. A second major earthquake struck Emilia-Romagna in 2012 during our sample period; Column 3 confirms that our estimates are robust to dropping households subsequently affected by that event.

Finally, the MEL cutoffs change across waves, so midpoint coding may assign different values to the same underlying discount factor in different waves. We run two checks (Table~\ref{tab:bincoding}). First, because the true discount factor lies within the reported bracket, its conditional mean is bounded by those of the bracket's lower and upper bounds \citep{manski2002}. Re-estimating the baseline with each respondent's lower or upper bound yields declines of 4.8 and 6.3 percentage points against the rest of Italy and 5.5 and 7.3 against the spatial control group, bracketing our baseline estimates of 5.5 and 6.4 (Columns 1--3). Since a difference-in-differences combines such means with opposite signs, this is a sensitivity check rather than a bound on the effect. Second, an interval regression \citep{harrison2002} assigns no value to an answer; it uses the probability that a normally distributed latent discount factor lies within the bracket offered in that wave. Its coefficients describe the latent discount factor itself, whose scale does not depend on where the cutoffs are drawn, so changing bins affect only how informative each answer is, not what the coefficients measure. As household fixed effects cannot be consistently estimated in this nonlinear model \citep{lancaster2000}, identification relies on cross-sectional variation, and the estimates are slightly larger than the OLS estimates in Table~\ref{tab:nakamura}. Across all these robustness checks, stated patience is declining in the treated group.

\section{Conclusion}\label{s:concl}
We propose a simple sign-restriction framework to identify
time-preference shocks: a genuine shock moves stated patience and
consumption in opposite directions.  A genuine impatience shock and a tightened financial constraint both lower stated patience, but imply opposite movements in consumption: the former raises it, the latter reduces it. Applying this framework to the 2009
L'Aquila earthquake, we combine stated time preferences, consumption, and
granular data on seismic exposure and damage.

Households that experienced the earthquake without damage became more
impatient than households that experienced no shaking: their stated
discount factor fell by five percentage points, nondurable consumption rose
by 11 percent, and saving remained lower for seven years.
To net out
spillovers and general equilibrium effects, we exploit the earthquake's
spatial spread and compare treated households with a spatial control group
located at a similar distance from the epicenter and the damaged area. This
group experienced a similar post-earthquake income decline, indicating
similar exposure to the earthquake's indirect effects.

Natural disasters can thus harm households even without physical damage,
by changing their choices. Our setting, a developed economy with public
reconstruction, universal health care, and developed credit markets, is
one where such effects might be expected to fade quickly. Yet treated
households save less for years, leaving them more exposed to future
shocks. Our results apply to unconstrained households and to a single
earthquake; as granular data become available for developing economies,
where insurance and credit markets are weaker and psychological and
material losses compound, similar mechanisms could produce larger and more
persistent effects. Disaster-risk assessments that count only material
destruction may therefore understate the economic exposure of affected
populations.

\bibliographystyle{ecta-fullname}%
\bibliography{references}%

\begin{appendix}
\section{Stated Preference Data} \label{appx:data}
\subsection{Discount factor}\label{appx:df}
The SHIW elicits time preferences from household heads in four waves (2004, 2008, 2010, and 2012) using the following money earlier or later question (MEL):
\begin{quote}
    Imagine that you find yourself in the following situation. You have won the lottery and will receive a sum equal to your household’s net yearly revenue. You will receive the
money in a year's time. However, if you give up part of the sum you can collect the rest of your win immediately.
\end{quote}
\noindent Respondents are then asked, in a stepwise manner, the \% of the win that they would give up to receive the rest immediately.%
To avoid inconsistent answers, if someone stated that she was willing to give up $2\%$ but not $5\%$, no further cutoffs were asked.\footnote{This question's simplicity yields a high response rate ($88\%$ on average). Recent experimental measures elicit across multiple time horizons, alongside quantitative risk preferences \citet{AndersenHarrisonLauRutstrom_ECTA08_risk_time}.} 

\subsection{Risk Aversion}
We measure risk preferences using the following investment attitude question, asked in every wave since 2004.\footnote{In 2004, only half of the population was asked this question.} 
\begin{quote}
    In managing your financial investments, would you say you have a preference for investments that offer: (1) very high returns, but with a high risk of losing part of the capital, (2) a good return, but also a fair degree of protection for the invested capital, (3) a fair return, with a good degree of protection for the invested capital, (4) low returns, with no risk of losing the invested capital.
\end{quote}
\noindent  In 2010, an Arrow-Pratt measure of risk aversion was additionally elicited via a lottery question:
\begin{quote}
    ``Imagine you can take part in a lottery in which for every euro invested, you can either double your money (win 1
euro) or lose half of it (lose 50 cents), depending on a coin toss (tails you win, heads you lose). How much money
would you invest?''
\end{quote}
Using this measure we impute a cardinal risk aversion measure in all waves, and correct the stated time preference measure for the curvature of the utility function (see details Appendix \ref{appendix:pref}).%

\begin{figure}[h!]
\centering
 \includegraphics[width=0.75\linewidth]{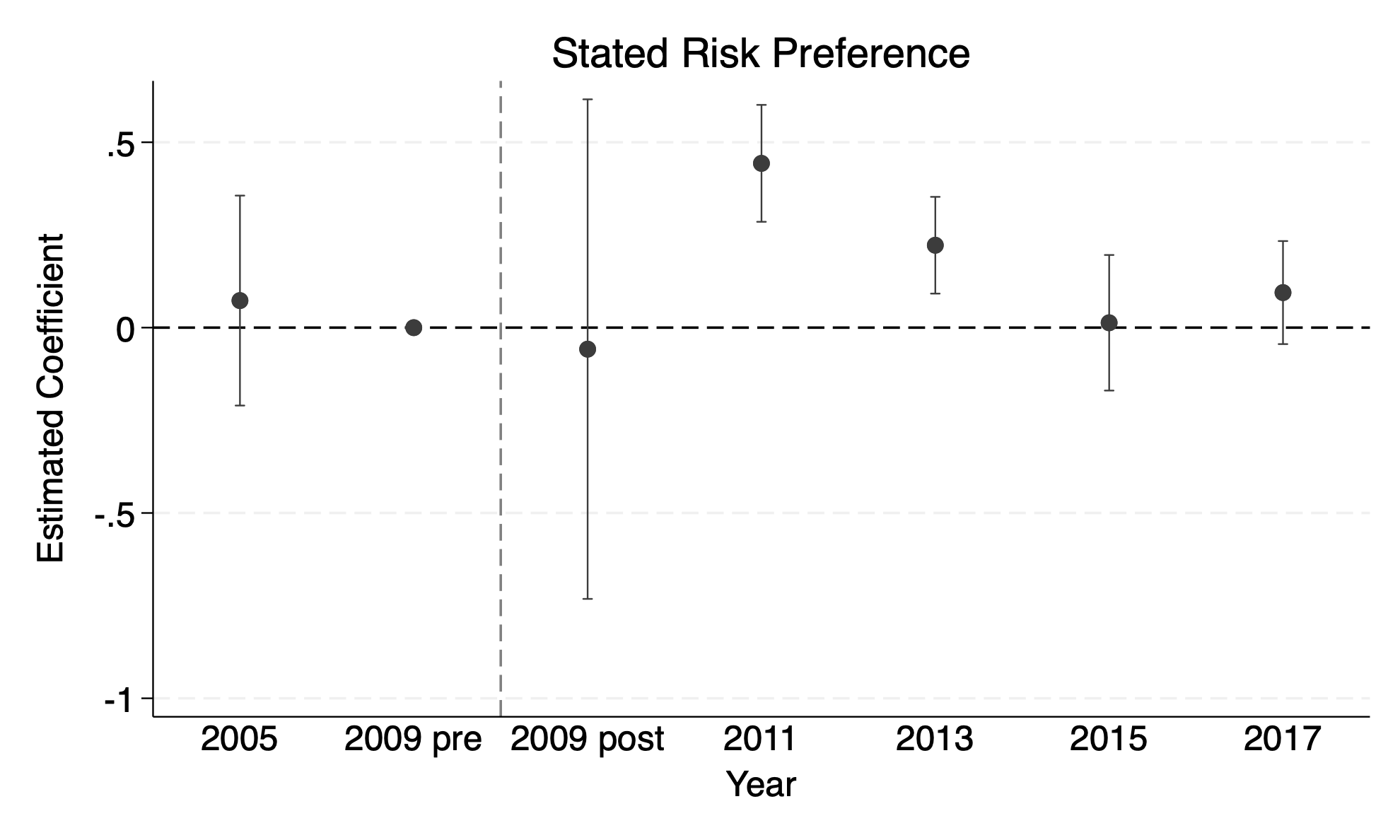}
         \caption{Stated Risk Preference. \textit{Notes}: Event study estimates of equation \ref{eq:es} (TWFE) on risk aversion. Measure:  a categorical variable 1-4 (1=most risk-loving, 4=most risk averse). Positive values indicate higher risk aversion. Reference period: 2008 wave, with interview dates before 6 April, 2009.}
    \label{fig:event_rp}
\end{figure}

\end{appendix}

\end{document}